\documentclass[12pt]{elsarticle}

\usepackage[english]{babel}
\usepackage[mac]{inputenc}
\usepackage{graphicx}
\usepackage{amssymb,amsfonts,amsmath,wrapfig,mathrsfs}
\usepackage{color}
\usepackage{epic} 
\usepackage{eepic} 
\usepackage{psfrag}
\usepackage{float}

\usepackage{verbatim}
\newcommand{\blue}{\color[rgb]{0,0,0}}

\journal{Applied Mathematical Modelling}

\begin{document}

\begin{frontmatter}



\title{Modelling the role of flux density and coating on nanoparticle internalization by tumor cells under centrifugation}

\author[1]{Gabriel F. Calvo$^*$}
\author[2,3]{Bel\'{e}n Cort\'{e}s-Llanos}
\author[1]{Juan Belmonte-Beitia}
\author[3]{Gorka Salas}
\author[3,4,5]{\'{A}ngel Ayuso-Sacido}

\address[1]{Department of Mathematics \& MOLAB-Mathematical Oncology Laboratory, Universidad de Castilla-La Mancha, Ciudad Real, Spain.}
\address[2]{Joint Department of Biomedical Engineering, University of North Carolina and North Carolina State University, Chapel Hill/Raleigh, NC  27599/27607, USA.}
\address[3]{IMDEA Nanoscience, Instituto Madrile\~{n}o de Estudios Avanzados, Madrid, Spain.}
\address[4]{HM Hospitales, Fundaci\'{o}n de Investigaci\'{o}n HM Hospitales, Madrid, Spain.}
\address[5]{Universidad San Pablo-CEU, Facultad de Medicina (IMMA), Madrid, Spain}


\cortext[Gabriel F. Calvo]{Contact email: {\em gabriel.fernandez@uclm.es}}
\cortext[Gabriel F. Calvo]{Manuscript published in {\em Applied Mathematical Modelling} {\bf 78}, 98-116 (2020).}




\begin{abstract}
Nanoparticle (NP)-based applications are becoming increasingly important in the biomedical field. However, understanding the interactions of NPs with biofluids and cells is a major issue in order to develop novel approaches aimed at boosting their internalization and, therefore, their translation into the clinic. 
To this end, we put forward a transport mathematical model to describe the spatio-temporal dynamics of iron oxide NPs and {\blue their} interaction with cells under moderate centrifugation. Our numerical simulations allowed us to quantify the relevance of the flux density as one of the unavoidable key features driving NPs interaction with the media as well as for cell internalization processes. These findings will help to increase the efficiency of cell {\blue labelling} for biomedical applications.
\end{abstract}


\end{frontmatter}

\section{Introduction}

The use of NPs in biological and medical areas demands understanding their interaction with media and cells at the nano-level scale \cite{Garcia15,Payne,Safi,Liz-Marzan}. Elucidating the involved processes in NPs-based systems is crucial in applications spanning from regenerative medicine to the diagnosis and treatment of a number of diseases such as cancer and ischemic stroke \cite{Edmunson13,Weissleder14,Bersen15,Betzer15,Gonzalez-Bejar16,Min17}. Iron oxide nanoparticles (IONPs) stand out among NPs because  they constitute a versatile vehicle to target cells, can be synthesized in a reproducible and scalable manner, delivered non-invasively and have been shown to be relatively safe and effective in different imaging modalities such as magnetic resonance imaging, radionuclide imaging, single-photon emission computed tomography \cite{Gao15,Iv15,Hachani17}. However, increasing the intracellular internalization in a significantly faster and controlled way remains one of the major challenges to fully translate and exploit their capabilities in the clinical practice. 
\par

Recently, the centrifugation-mediated internalization method has emerged as an effective alternative to direct incubation methods as it meets the required demands~\cite{Ocampo}. Nevertheless, the key underlying processes driving the delivery and intracellular {\blue uptake} of IONPs in these methods still need to be explained. {\em In silico} approaches provide powerful tools to access those processes which may be technically demanding to measure {\em in vitro} or {\em in vivo}. One specific framework that has been successfully applied to describe bionano experiments is based on employing diffusion-sedimentation evolution equations~\cite{Hinderliter,Stellacci,Cui,Bekdemir}. 
\par

These computational models require as input data characteristics of the IONPs such as their geometric/hydrodynamic radio and density or their concentration at various time frames, whose availability becomes an important limitation when facing real systems involving complex NPs-coatings and interactions with cellular subsystems for sustained periods of time. These scenarios make necessary the development of new {\blue modelling} strategies. Indeed, this is of special relevance for IONPs dispersed in a cell culture medium, where they experience forces with proteins and other molecules as well as with cell membranes. Eventually, these processes may lead to the formation of the so-called protein corona~\cite{Mahmoudi,monopoli2012biomolecular,lundqvist2008nanoparticle}, in which some proteins are adsorbed or linked to the IONPs surface. Aggregation is another typical process that can occur when IONPs are dispersed in cell culture media~\cite{Moore,russel1989colloidal}. This may be due to changes in the ionic strength and the subsequent shielding of the surface charge or because of the substitution of the coating by molecules and macromolecules present in the medium. Both processes can totally or partially destabilize the IONPs dispersion. In addition to these processes, elucidating the cellular internalization of the IONPs and identifying relevant temporal metrics entails for the diffusion-sedimentation framework to be substantially extended.
\par

\begin{figure}[t]
\centering
\includegraphics[width=0.7\linewidth]{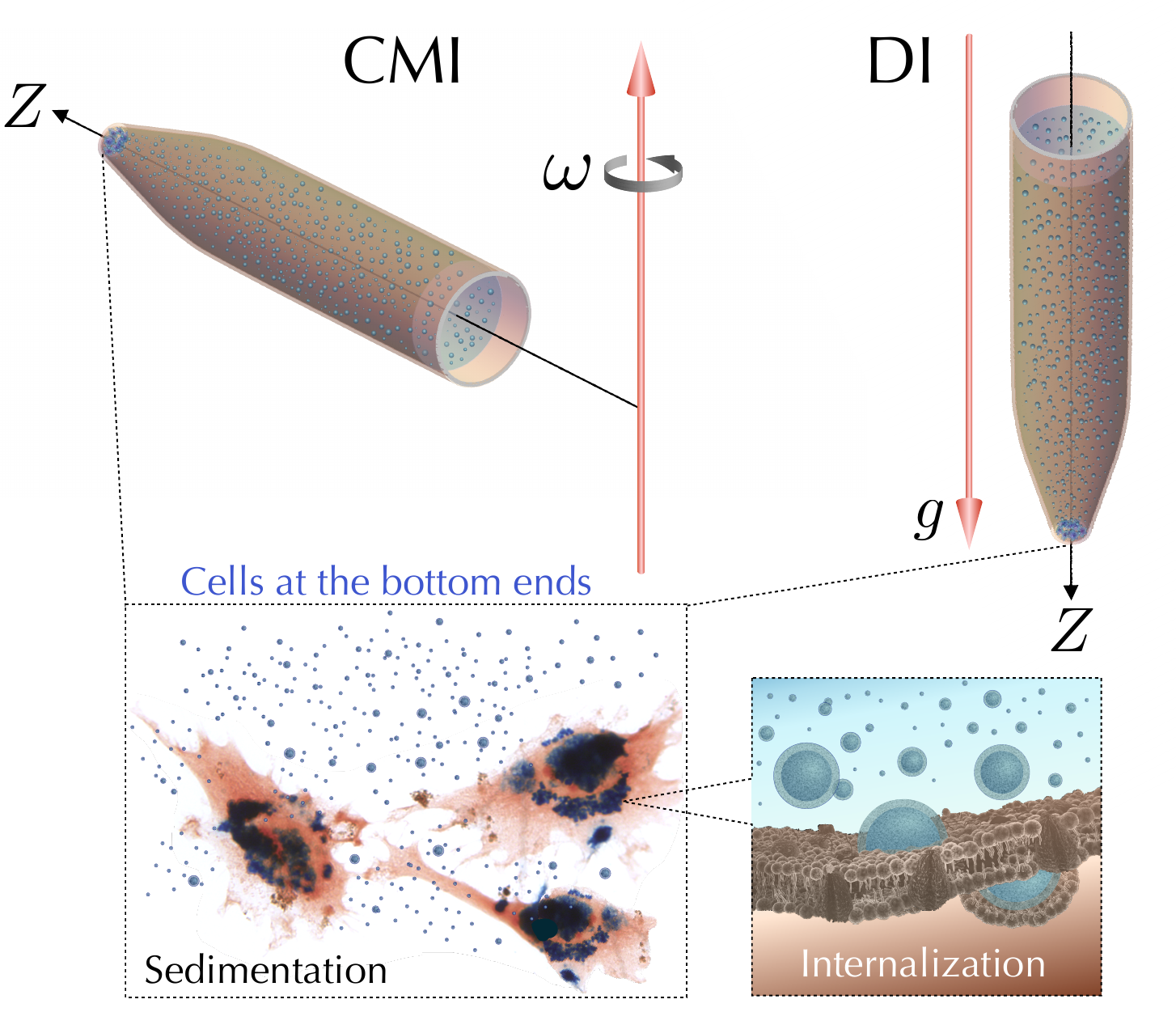}
\caption{Schemes of centrifugation-mediated internalization (CMI) and direct incubation (DI) configurations. In both cases a cylindrically-symmetric container comprises the solvent, the IONPs and the cells pelleted at the {\blue bottom (narrow)} ends. Once {\blue the IONPs reach the bottom ends of the containers, they} display adsorption, desorption and internalization processes within the cell membrane vicinity {\blue (lower-right inset)}.}
\label{fig:GeometryCentrifGrav}
\end{figure} 

To induce cellular internalization of IONPs, two major {\em in vitro} configurations can be distinguished; direct or gravity-mediated incubation (DI) and CMI (Figure \ref{fig:GeometryCentrifGrav}). In DI, IONPs sediment due to the sole action of gravity, whereas cells are typically placed at the bottom of the culture medium, although other orientations are also worth of interest (e.g. lateral and upside down). Under CMI, IONPs move through the medium due to the action of large centrifugal forces towards cells pelleted at the bottom end of the rotating container, while a competition between two opposing mechanisms arises: sedimentation and diffusion. The non inertial centrifugal force produces a collective motion of the IONPs towards one end of the container where they sediment. This in turn creates a concentration gradient of the IONPs leading to a diffusion flux. Thus, the total flux of IONPs is the result of sedimentation and diffusion fluxes and plays a key role to completely quantify IONPs internalization by cells. 
\par

To gain insight into these processes, we used a set of IONPs with the same inorganic core and five distinct coatings under {\blue the} centrifugation-mediated internalization (CMI) method. We studied their {\blue hydrodynamic} diameter by dynamic light scattering measurements (DLS). Then, we {\blue proceeded to analyse} the cellular internalization of IONPs, by considering two {\em in vitro} configurations: DI, and CMI \cite{Ocampo}. To further quantify our observations, both in the absence and in the presence of cells, and under centrifugal/gravitational fields, we developed a unified kinetic transport modeling framework to encompass all these scenarios.
\par

\section{Experimental procedure}

The internalization of IONPs into cells is strongly influenced by their coating and colloidal properties~\cite{saei2017nanoparticle}. In order to evaluate the role of the different coatings, we have used the same IONP core with five different coatings to assess their CMI-mediated internalization capabilities. IONPs with a mean core size of  $14.4\pm 3.7$ nm were prepared by the classical Massart's method~\cite{massart1981preparation}, followed by an oxidative acid treatment to obtain maghemite NPs \cite{Costo}. The IONPs with different coatings were named as follows (coating in brackets): NP (naked), NP-D (dextran), NP-AD (amino-dextran), NP-CMD (carboxymethyl-dextran), NP-APS (aminopropyl-trietoxy silane), NP-DMSA (dimercaptosuccinic acid). When NPs are dispersed in a liquid medium they frequently undergo aggregation~\cite{Moore,russel1989colloidal}. Aggregation can be studied via the DLS technique and will influence the interaction of IONPs with the cellular membrane. Dextran derivatives (NP-D, NP-AD and NP-CMD) provide steric stabilization, while using NP, NP-DMSA or NP-APS, there is no steric hindrance and the stabilization is due to the electrostatic repulsion. We resuspended the IONPs in dulbecco's modified Eagle's medium (DMEM). This is a synthetic cell culture medium widely used to maintain cells in tissue cultures. We supplemented DMEM with 10$\%$ of FBS or w/o FBS to reach the best administration of IONPs and to avoid the formation of agglomerates by CMI methodology. Prussian blue images show that using NP-DMSA resuspended on DMEM with 10$\%$ of FBS the highest labelling efficiency was observed with a {\blue hydrodynamic} diameter of 335 nm. 
\par

Once the IONPs were synthesized and characterized,  we applied the CMI (without cells) to determine their spatio-temporal distribution in the rotating tube. This would enable us to estimate the sedimentation and diffusion coefficients corresponding to each IONP type. The used media containing the suspended IONPs was Dulbecco's modified eagle's medium (DMEM) or fetal bovine serum (FBS). The initial IONPs concentration was 50 $\mu$g/ml in 5 ml. After the CMI method (1500 rpm for 5 min), the final volume (5 ml) was fractionated in volumes of 1 ml (four times), 200 $\mu$l (four times) and 100 $\mu$l (two times). The measurements were made at $t=0$ h (immediately after CMI), after 3 h and 10 h later on (see also Subsection 3.2 for additional details). By using Prussian blue staining we measured the Fe absorbance at 690 nm employing a UV-visible spectrophotometer (Sinergy H4 microplate reader). For the IONPs quantification, two different FeCl$_{3}$ calibration curves were performed, one in DMEM and another one in DMEM supplemented with 10$\%$ FBS. Finally, each IONPs concentration was normalized to the total one. 
\par
A U373 glioblastoma cell line was obtained from the American Type Culture Collections (Manassas, VA, USA). The cell line was grown in DMEM supplemented with $10\%$ of FBS, $2$ mM of L-glutamine,  1 $\mu$g/ml of fungizone, 100 $\mu$g/ml streptomycin and $100$ unit of penicillin per ml (GIBCO). Cell line was maintained at 37$^{\circ}$C in a humidified atmosphere of $95\%$ air and $5\%$ CO$_{2}$.

\section{Mathematical procedure}

\subsection{Models of IONPs sedimentation and diffusion}

Let $c(z,t):[z_{a},z_{b}]\times(0,T)\to\mathbb{R}^{+}$ denote the IONPs mass concentration distribution where $z\in[z_{a},z_{b}]$ denotes the position along the longitudinal axis of a cylindrical container, and $t\in(0,T)$ is time. Here $z_{a}>0$ and $z_{b}>0$ denote the positions of the meniscus and the {\blue bottom end of the container (see Figure \ref{fig:GeometryCentrifGrav})}, respectively, and $T>0$ is the time duration considered. The evolution of $c(z,t)$, in any of the two studied configurations CMI and DI, is governed by the following continuity equation 
\begin{eqnarray}
\frac{\partial c}{\partial t} = - \frac{\partial J}{\partial z}\, ,
\label{eq:Continuity}
\end{eqnarray}
where $J$ denotes the IONP flux density {\blue along the $Z$ axis}. Equation (\ref{eq:Continuity}) expresses the conservation of the total mass of IONPs within the solvent since {\blue IONPs are neither} created nor destroyed during the experiments. Depending on the used configuration, the flux density is 
\begin{eqnarray}
J=J_\textrm{L}\equiv s\omega^{2}zc - D\frac{\partial c}{\partial z}\, ,
\label{eq:JL}
\end{eqnarray}
when a centrifugation force is applied, with $\omega>0$ being a constant angular velocity (the longitudinal symmetry $Z$ axis of the container is positioned perpendicularly to the rotation axis, see {\blue left side in} Fig.~\ref{fig:GeometryCentrifGrav}). Here, $D>0$ and $s>0$ represent the diffusion and the sedimentation coefficients, respectively.
\par
Under the sole presence of the acceleration due to gravity $g$ (directed downwards along the $Z$ axis, see {\blue right side in} Fig.~\ref{fig:GeometryCentrifGrav}), the flux density is 
\begin{eqnarray}
J={\blue J_\textrm{G}}\equiv {\blue sgc} - D\frac{\partial c}{\partial z}\, .
\label{eq:JG}
\end{eqnarray}
\par

As shown in the Appendix, the partial differential equation obeyed by the IONPs concentration $c(z,t)$ under centrifugation is 
\begin{eqnarray}
\frac{\partial c}{\partial t} = D\frac{\partial^{2} c}{\partial z^{2}} -s\omega^{2}\left( z\frac{\partial c}{\partial z} +c\right)\! ,
\label{eq:LammCylZ}
\end{eqnarray}  
that holds for $z\in[z_{a},z_{b}]$ and $t>0$. The imposed boundary conditions (of Neumann type) are
\begin{eqnarray}
J(z_{a},t)=0 \, , \quad J(z_{b},t)=0\, ,
\label{eq:BC}
\end{eqnarray}
while the initial condition is $c(z,0) = c_{0}(z):[z_{a},z_{b}]\to\mathbb{R}^{+}$. 
\par

We wish to underscore that equation (\ref{eq:LammCylZ}) is similar in form, but not identical, to the standard Lamm equation~\cite{Demeler,Schuck} for a radially symmetric geometry, which was not the case in our experimental setting, although both equations encompass diffusive and sedimentation mechanisms and predict similar phenomena. The standard Lamm equation has been previously applied to describe bionano experiments~\cite{Stellacci,Bekdemir}. Depending on the value of the angular velocity $\omega$, two regimes can be identified: sedimentation velocity and sedimentation equilibrium. In the former one, analytes are centrifuged under very high angular speeds ($\omega > 10^{4}$ rpm) and become completely separated (precipitated) from the solvent. In the second regime, lower $\omega$'s are employed which makes the role of diffusion comparatively more important. This last scenario is the one considered in our experiments as $\omega = 1500$ rpm.

\par
When gravity is the only external force acting on the IONPs, sedimentation can also take place although at a significantly slower pace when compared with the application of centrifugation. In our experimental setting, the cylindrical container, comprising the solvent and the IONPs, {\blue was then} placed with its longitudinal symmetry axis along the gravity field (i.e. the $Z$ axis), as shown on the right of Fig. \ref{fig:GeometryCentrifGrav}). In the Appendix we provide the detailed derivation of the partial differential equation that describes the dynamics of $c(z,t)$ under the sole action of gravitation-mediated sedimentation and Fickian diffusion. This equation reads as
\begin{eqnarray}
\frac{\partial c}{\partial t} = D\frac{\partial^{2} c}{\partial z^{2}} -sg\frac{\partial c}{\partial z} \, ,
\label{eq:PDEGravity}
\end{eqnarray} 
where $z\in[z_{a},z_{b}]$ and $t>0$, with $c(z,t)$ satisfying initial condition $c(z,0) = c_{0}(z)$ and boundary conditions (\ref{eq:BC}) {\blue with the flux density given by (\ref{eq:JG})}.
\par

To describe the spatio-temporal dynamics of $c(z,t)$ {\blue with different IONPs} and to predict {\blue the behaviour} depending on whether {\blue the} DI or CMI configurations {\blue are} employed, we first examine the scenario of IONPs suspension in the absence of cells for {\blue these} two internalization methodologies. To gain insight about the sedimentation process of the IONPs, we performed intensive numerical simulations by means of an adaptive finite element method~\cite{Cao} to solve Eqs.~(\ref{eq:LammCylZ}) and (\ref{eq:PDEGravity}) using different values of the sedimentation and diffusion coefficients. An important difference between the corresponding equations, either under gravity or centrifugation, is that, in the first case, an initially uniform concentration distribution does not lead to a nonuniform profile for $t>0$ as it occurs in the second one. In practice, the presence of small perturbations in an otherwise quasi-uniform initial concentration distribution is sufficient to observe sedimentation mediated by gravity.
\par

\begin{figure}[h!]
\centering
\hspace*{-2mm}
\includegraphics[width=0.99\linewidth]{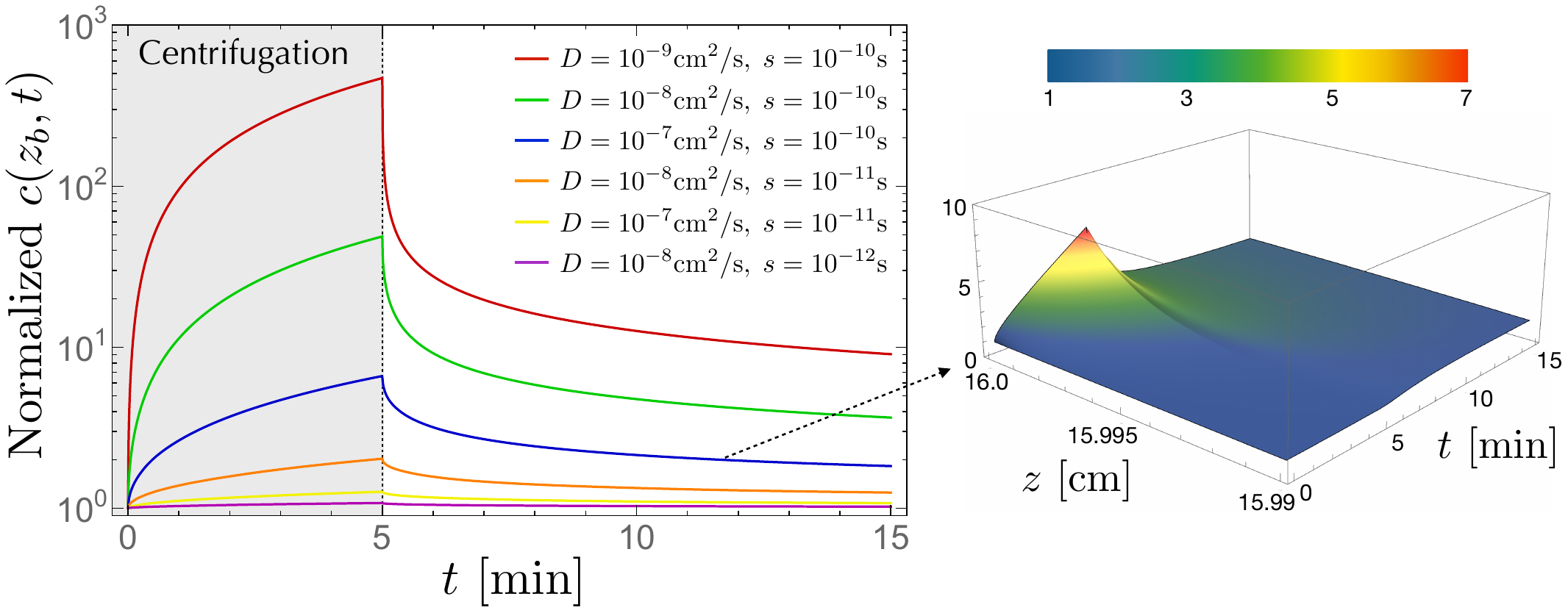}
\caption{\small \blue Temporal profiles of the mass concentration of IONPs at the bottom end ($z=z_{b}=16$ cm) of the container under successive application of centrifugation (with $\omega = 1500$ rpm) during 5 minutes followed by gravity sedimentation for a total time of 10 minutes. The profiles $c(z_{b},t)$ are calculated from Eqs. (\ref{eq:LammCylZ}) and (\ref{eq:PDEGravity}) for different values of the diffusion and sedimentation coefficients and are normalized with respect to an initial uniform concentration. Numerically, at $z=z_{b}$, the temporal profiles were virtually undistinguishable when $\frac{s^{2}}{D}$ attained the same ratio. The 3D plot on the right displays the spatio-temporal profile of the mass concentration in the vicinity of the bottom end of the container for $D=10^{-7}$ cm$^{2}$/s and $s = 10^{-10}$ s.}
\label{fig:LammGrav}
\end{figure}

{\blue Figure \ref{fig:LammGrav} summarizes the temporal profiles of the IONPs in the vicinity of the bottom end of the cylindrical container during centrifugation-mediated sedimentation for 5 minutes, with a constant angular velocity $\omega = 1500$ rpm, followed by a gravitation-mediated sedimentation interval of 10 minutes. In all cases, we launched the simulations assuming a uniform initial condition $c(z,t=0) = c_{0}$ for $z\in[z_{a},z_{b}]$, with $c_{0}$ a constant value.} The solvent occupied a length interval from $z_{a}=11$ cm (the meniscus) to $z_{b}=16$ cm ({\blue bottom end}). During the centrifugation phase, our simulations showed that a narrow {\blue (below 100 $\mu$m thickness)} boundary layer in the IONP concentration developed in the proximity of the {\blue bottom end} of the cylindrical container. This layer is a consequence of the IONP flux towards the bottom of the container, where significant IONP-wall collisions and IONP accumulation take place. Its amplitude and thickness depends on the specific values of the sedimentation $s$ and diffusion $D$ coefficients. The ranges considered in Figure~\ref{fig:LammGrav} were $s\in[10^{-12},10^{-10}]$ s and $D\in[10^{-9},10^{-7}]$ cm$^{2}$/s. {\blue The amplitude of the IONP concentration layer increased either with centrifugation time or with the ratio $\frac{s^{2}}{D}$.} Far from the {\blue bottom end, $c(z,t)$ exhibited an uniform plateau up to the meniscus (see the 3D plot on the right of Figure \ref{fig:LammGrav})}.
\par

\subsection{\blue Determination of the sedimentation and diffusion coefficients for IONPs}

\begin{figure}[h!]
\centering
\hspace*{-1.5mm}
\includegraphics[width=1.01\linewidth]{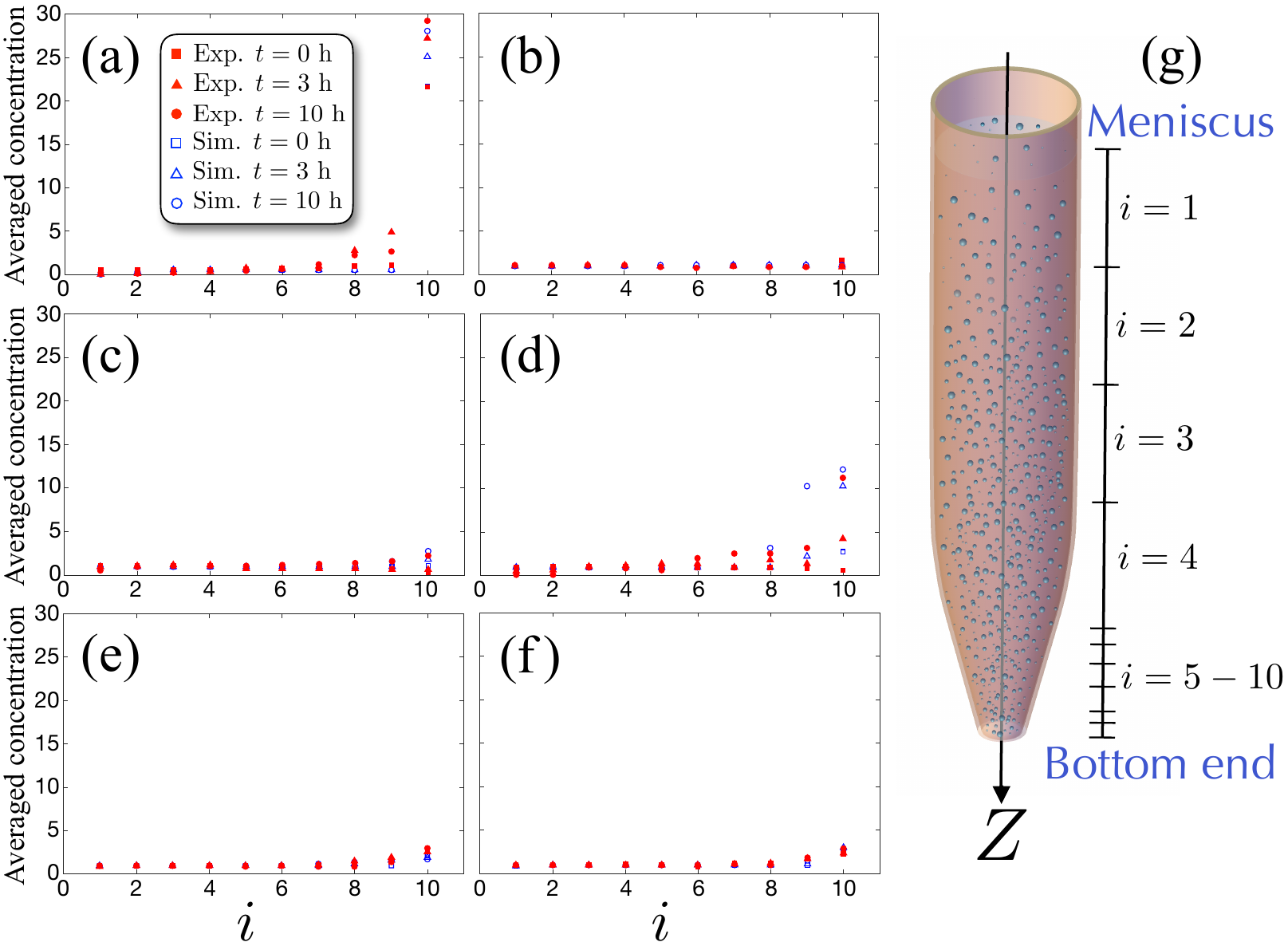}
\caption{Averaged concentrations measured and simulated for: (a) NP; (b) NP-D; (c) NP-AD; (d) NP-CMD; (e) NP-APS; and (f) NP-DMSA IONPs under the successive application of centrifugation (with $\omega=1500$ rpm) for 5 minutes followed by gravitation-mediated sedimentation at times: 0, 3, and 10 hours. The abscissae $i$ correspond to a sequence of subvolumes sampled along the cylindrical container (total volume of 5 ml) starting from the meniscus down to the {\blue bottom end}: $i=$ 1-4 stand for 1 ml subvolumes, $i=$ 5-8 for 200 $\mu$l subvolumes and $i=$ 9-10 for 100 $\mu$l subvolumes. {\blue (g) Schematic of the volume samples taken from the container in the experiments to measure the averaged concentrations of all the IONPs at different distances.}}
\label{Fe-NPs_exp_teo}
\end{figure}

{\blue 

To understand how the IONPs reach the region were cells are pelleted, we first measured the IONPs stratification along the solution media by quantifying them at different volume sections within the cylindrical container as explained in Section 2. The experimental procedure consisted of applying centrifugation (with $\omega = 1500$ rpm) to the cylindrical container during 5 minutes followed by gravity sedimentation at various time intervals of 0, 3 and 10 hours. Samples were collected from the meniscus down to the bottom end of the container. The total volume of the solution contained in the tube was 5 ml. Since the concentration $c(z,t)$ profiles were not directly accessible during the experiments, to assess the values of the sedimentation and diffusion coefficients corresponding to the different IONPs studied, we reproduced {\em in silico} the measurement of the averaged IONP concentrations, defined as
\begin{eqnarray}
\langle c_{i} (t) \rangle = \frac{1}{z_{i+1} -z_{i}}\int_{z_{i}}^{z_{i+1}} c(z,t) dz ,
\label{eq:AverageConcentration}
\end{eqnarray} 
where $z_{i}$, with $z_{a}\leq z_{i}\leq z_{b}$, denotes the position along the $Z$ axis of the cylindrical container of a disk-like subvolume $\Delta V_{i}$ of section $\pi R_{i}^{2}$ and thickness $\Delta z_{i}=z_{i+1} -z_{i}$. The averaged IONPs concentrations $\langle c_{i} (t) \rangle$ were calculated from Eqs.~(\ref{eq:LammCylZ}), (\ref{eq:PDEGravity}) and (\ref{eq:AverageConcentration}). To mimic the experimental conditions, we computed the concentration $c(z,t)$ under centrifugation-mediated sedimentation for 5 minutes, with a constant angular velocity $\omega = 1500$ rpm, followed by a gravitation-mediated sedimentation interval of up to 10 hours. In all cases, we launched the simulations assuming a uniform initial condition $c(z,t=0) = c_{0}$ for $z\in[z_{a},z_{b}]$, with $c_{0}$ being a constant value.
\par

Figure \ref{Fe-NPs_exp_teo}(a)-(f) displays a comparison between our experimental and numerical simulation results for the averaged concentrations of all the IONPs studied. The index $i$ runs from $i=1$-10, starting from the meniscus down to the {\blue bottom end [see Fig.~\ref{Fe-NPs_exp_teo}(g)]}. They correspond to $i=$1-4 for 1 ml subvolumes, $i=$5-8 for 200 $\mu$l subvolumes and $i=$9-10 for 100 $\mu$l subvolumes.  In the simulations the sedimentation and diffusion constants were first estimated from previously measured hydrodynamic diameters, probing a broad range of values, and then fitted. The listed values of the effective $s$ and $D$ found for all the six IONPs are provided in Table \ref{table_Parameters}. 
\par

}

\begin{table}
\begin{center}
\begin{tabular}{|c|c|c|}%
\hline
 IONPs & $s$ [s] & $D$ [cm$^2/$s] \\
\hline
NP & $1.0\cdot 10^{-10}$ & $5.0\cdot 10^{-9}$ \\
NP-D & $1.0\cdot 10^{-10}$ & $1.2\cdot 10^{-8}$ \\
NP-CMD  & $6.0\cdot 10^{-9}$ & $9.0\cdot 10^{-8}$ \\
NP-AD & $3.0\cdot 10^{-9}$  & $2.0\cdot 10^{-9}$ \\
NP-APS & $2.0\cdot 10^{-9}$ & $7.0\cdot 10^{-7}$ \\
NP-DMSA  & $5.0\cdot 10^{-10}$ & $1.1\cdot 10^{-7}$\\
\hline
\end{tabular}
\caption{Sedimentation and diffusion parameters for each IONP that were used in the simulations of Eq.~(\ref{eq:PDEGravity}) to fit the experimental results shown in {\blue Fig.~\ref{Fe-NPs_exp_teo}}.}
\label{table_Parameters}
\end{center}
\end{table}

In all the IONPs that we measured, the spatial dependence of the concentration displayed an exponential profile progressively increasing towards the bottom end of the container, in agreement with our calculations. Immediately after centrifugation ($t=0$ hours), the concentration at the bottom end reached a maximum value only for NP-D, which very slightly decreased with time [see Figure~\ref{Fe-NPs_exp_teo}(b)]. In contrast, all other IONPs exhibited a variable growth in the concentration at the bottom end during the gravitation-mediated sedimentation phase (at $t=3$ and $t=10$ hours), which was most prominent for naked NP and NP-CMD [Figures ~\ref{Fe-NPs_exp_teo}(a) and (d)]. {\blue For naked NP, the absence of any protective coating gives rise to a smaller electrostatic stabilization; that is, to a much smaller intermolecular interaction of these NPs with the solvent medium. In this type of IONPs aggregation will be larger when compared with all other coated NPs. Furthermore, when dispersed in a cellular medium, the hydrodynamic diameter (related with the aggregate size) of naked NP is higher than in the other IONPS, producing less stability and agglomeration. This will result in a sustained sedimentation at the bottom end of the container, mediated by gravity, that will keep increasing with time (for at least the duration of our measurements). The behaviour shown by NP-AD and NP-CMD during the centrifugation phase [Figures~\ref{Fe-NPs_exp_teo}(c) and (d) at $t=0$ hours] can also be explained in terms of the steric interaction. In these two cases, there was an increase (less marked than for naked NP) of the spatial concentration at the bottom end during the gravitation-mediated sedimentation phase.  Comparatively, NP-APS and NP-DMSA [Figures~\ref{Fe-NPs_exp_teo}(e) and (f)] did not evidence a significant increase in the concentration at the bottom end during all the gravitation-mediated sedimentation phase, yet their concentrations at $t=0$ hours were systematically higher than for the dextran-derivate IONPs.} Consequently, this suggests that for dextran-derivate (CMD) IONPs the DI method would be preferred to the CMI method in order to achieve higher cellular uptake~\cite{Deng}. On the other hand, NP-APS and NP-DMSA seem to perform much better by the use of g-forces, which only requires a few minutes.  
\par

\subsection{Models of IONPs internalization by cells}

To elucidate the interaction of IONPs with cells, either under CMI or DI, we extended model (\ref{eq:Continuity}) to incorporate three additional processes: (i) adsorption (or binding) of the IONPs onto the cell membrane, (ii) desorption of the IONPs from the cell membrane, and (iii) internalization of the IONPs into the cytosol. To quantify these processes we put forward a Langmuir-type kinetic model that also includes the spatial distribution of cells via a density function $\rho_\textrm{cell}=\rho_\textrm{cell}(z)$. This localized profile, which is nonzero only in the vicinity of the {\blue bottom end} (in a region of width $60$-$80$ $\mu$m), is justified by the preparation method of the cell cultures. We also define two ancillary quantities, $c_\textrm{b}=c_\textrm{b}(z,t)$ and $c_\textrm{i}$ = $c_\textrm{i}(z,t)$, that represent the mass concentrations of bound and internalized IONPs, respectively. To take into account that these two concentrations are each limited by the maximum number of available binding sites at the cell membrane and by the carrying capacity of IONPs internalized by the cells, two saturation functions $S_\textrm{b}$ and $S_\textrm{i}$ were employed. 
\par

Under centrifugation, the equations that govern the cellular uptake of IONPs are
\begin{subequations}\label{eq:LammCells}
\begin{eqnarray}
\frac{\partial c}{\partial t} \!\! &=&\!\! -\frac{\partial J_\textrm{L}}{\partial z} + k_\textrm{d} c_\textrm{b} - \sigma_\textrm{b}S_\textrm{b}J_\textrm{L}\rho_\textrm{cell},
\label{eq:LammCells1} \\
\frac{\partial c_\textrm{b}}{\partial t} \!\! &=&\!\! \sigma_\textrm{b}S_\textrm{b}J_\textrm{L}\rho_\textrm{cell} - k_\textrm{d} c_\textrm{b} - k_\textrm{i}S_\textrm{i}c_\textrm{b},
\label{eq:LammCells2}\\
\frac{\partial c_\textrm{i}}{\partial t} \!\! &=&\!\!  k_\textrm{i}S_\textrm{i}c_\textrm{b}.
\label{eq:LammCells3}
\end{eqnarray}  
\end{subequations}
\noindent In (\ref{eq:LammCells}), a fraction of those IONPs that reach, by sedimentation, the bottom of the tube where the cells are located, bind to an available receptor on the plasma membrane. This is described by the first term on the right-hand-side of (\ref{eq:LammCells2}). Here, $\sigma_\textrm{b}$ is a binding constant (dimensionless) and $S_\textrm{b}=S_\textrm{b}\left(\frac{c_\textrm{b}}{c_\textrm{b}^\textrm{(max)}}\right)$ a dimensionless function to account for the saturation of the available cell membrane binding sites. The desorption process is represented by the second term on the right-hand-side of (\ref{eq:LammCells2}), where $k_\textrm{d}$ is a rate constant (with units of inverse time). Subsequently, the receptor-bound IONPs are internalized, as represented by the right-hand-side of (\ref{eq:LammCells3}), where $k_\textrm{i}$ is a rate constant (with units of inverse time) and $S_\textrm{i}=S_\textrm{i}\left(\frac{c_\textrm{i}}{c_\textrm{i}^\textrm{(max)}}\right)$ a dimensionless function to model the saturation effect of the IONPs uptake. These rate constants depend on the specific corona of the IONPs. The adsorption kinetics embodied in (\ref{eq:LammCells1}) and (\ref{eq:LammCells2}) considers that it is the flux density $J_\textrm{L}(z,t) = J_{z}= s\omega^{2}zc - D\frac{\partial c}{\partial z}$, rather than only the local concentration $c(z,t)$, the key underlying mechanism. The flux density $J_\textrm{L}$ would be proportional to $c(z,t)$ if $- D\frac{\partial c}{\partial z}$ were negligible, which occurs if either the concentration gradient or the diffusion coefficient are exceedingly small. Otherwise, there is a significant departure from a purely local dependence with the IONPs concentration. The larger the value of $J_\textrm{L}$ is at the cellular region (due to a larger sedimentation coefficient $s$ or to a larger angular velocity $\omega$ or both), the faster the adsorption process becomes. Furthermore, it is expected that $c_\textrm{b}^\textrm{(max)} < c_\textrm{i}^\textrm{(max)}$, based on the observation that the maximum mass concentration of bound IONPs onto the cell membrane is smaller than the maximum mass concentration of IONPs internalized by the cell, thus reflecting the fact that it is the mass of adsorbed IONPs the limiting process. 
\par
Under the sole action of gravity, the equations that govern the cellular uptake of IONPs are
\begin{subequations}\label{eq:GravCells}
\begin{eqnarray}
\frac{\partial c}{\partial t} \!\! &=&\!\!  -\frac{\partial J_\textrm{G}}{\partial z} + k_\textrm{d} c_\textrm{b} - \sigma_\textrm{b}S_\textrm{b}J_\textrm{G}\rho_\textrm{cell},
\label{eq:GravCells1} \\
\frac{\partial c_\textrm{b}}{\partial t} \!\! &=&\!\!  \sigma_\textrm{b}S_\textrm{b}J_\textrm{G}\rho_\textrm{cell} - k_\textrm{d} c_\textrm{b} - k_\textrm{i}S_\textrm{i}c_\textrm{b},
\label{eq:GravCells2}\\
\frac{\partial c_\textrm{i}}{\partial t} \!\! &=&\!\!  k_\textrm{i}S_\textrm{i}c_\textrm{b}.
\label{eq:GravCells3}
\end{eqnarray}  
\end{subequations}

The interpretation of the adsorption, desorption and uptake processes in (\ref{eq:GravCells}) is analogous to (\ref{eq:LammCells}); the key difference is the absence of a strong IONP gradient concentration in the vicinity of the {\blue bottom end} of the container. As pointed out above, if the IONP concentration $c$ is uniform, then the first term in the right-hand-side of (\ref{eq:GravCells1}) vanishes. However, the presence of cells is expected to modify the concentration, even if it is initially uniform along the cylindrical container. Again, incorporating the flux density $J_\textrm{G}(z,t)$, and not just a local concentration $c(z,t)$, drives the adsorption kinetics, although at a significantly slower pace in comparison with CMI since $J_\textrm{G}(z,t)\ll J_\textrm{L}(z,t)$, for the same $s$ and $D$ coefficients. Indeed, under gravity alone, the flux density $J=J_\textrm{G}$ is approximately proportional to $c(z,t)$, due to the small correction by diffusion which is often disregarded, even in the vicinity of the cellular region. This justifies why previously used Langmuir-type kinetic models did not consider the role played by the IONPs flux when describing cellular uptake. However, under centrifugation, the diffusive term in the flux density $J=J_\textrm{L}$ cannot be neglected in the cellular region, at the bottom end of the rotating container. In fact, the larger the value of $J_\textrm{L}$ is at the cellular region, due to a larger product $s\omega^{2}$ and/or to the existence of strong negative gradients in the concentration, the faster the adsorption process becomes. Therefore, Langmuir-type kinetic models should incorporate the action of the IONPs flux density to provide an accurate description of the interplay between IONPs and the cellular membranes. 
\par

\begin{table}[h!]
\caption{General parameters used in all the simulations}
\label{tab:GlobalParameters}
\vspace*{1mm}
\begin{center}
\begin{tabular}{|c|c|c|}
\hline
 Symbol & Description & Range of values  \\
\hline
$z_{a}$ & Position of the meniscus & $11$ cm \\
$z_{b}$ & Position of the {\blue bottom end} & $16$ cm \\
$\omega$ & Angular velocity & 1500 rpm\\
$g$ & Acceleration due to gravity & $981$ cm/s$^{2}$ \\
\hline
\end{tabular}
\end{center}
\end{table}
\par

The mass of IONPs internalized by the cells is calculated by means of 
\begin{eqnarray}
M_\textrm{i}(t) = \int_{z_{a}}^{z_{b}}c_\textrm{i}(z,t)\, dz,
\label{eq:InterMass}
\end{eqnarray}  
where $c_\textrm{i}(z,t)$ obeys either (\ref{eq:LammCells3}) or (\ref{eq:GravCells3}) depending on whether CMI or DI is considered. From $M_\textrm{i}(t)$, the IONP content per cell as a function of centrifugation time can be calculated if the number of pelleted cells is known. To gauge the relative importance of $M_\textrm{i}$, we also define the total mass $M_\textrm{tot}$ of IONPs contained in the solvent. Since IONPs are assumed not to be created nor destroyed during sedimentation, $M_\textrm{tot}$ is a constant.
\par

\begin{table}[b]
\caption{Kinetic parameters of NP-DMSA}
\label{tab:ParametersDMSA}
\vspace*{1mm}
\begin{tabular}{|c|c|c|}
\hline
 Symbol & Description & Range of values  \\
\hline
$k_\textrm{d}$ & Desorption rate constant & $0.47\times10^{-2}$ s$^{-1}$  \\
$k_\textrm{i}$ & Internalization rate constant &  $0.14$ s$^{-1}$  \\
$\sigma_\textrm{b}$ & Binding constant & $7.0\times10^{4}$  \\
$c_\textrm{b}^{(\textrm{max})}$ & Saturation of cell membrane concentration & 0.3 \\ 
$c_\textrm{i}^{(\textrm{max})}$ & Saturation of cell uptake concentration & 7.0 \\
\hline
\end{tabular}
\end{table}

\begin{figure}[t!]
\centering
\includegraphics[width=0.9\linewidth]{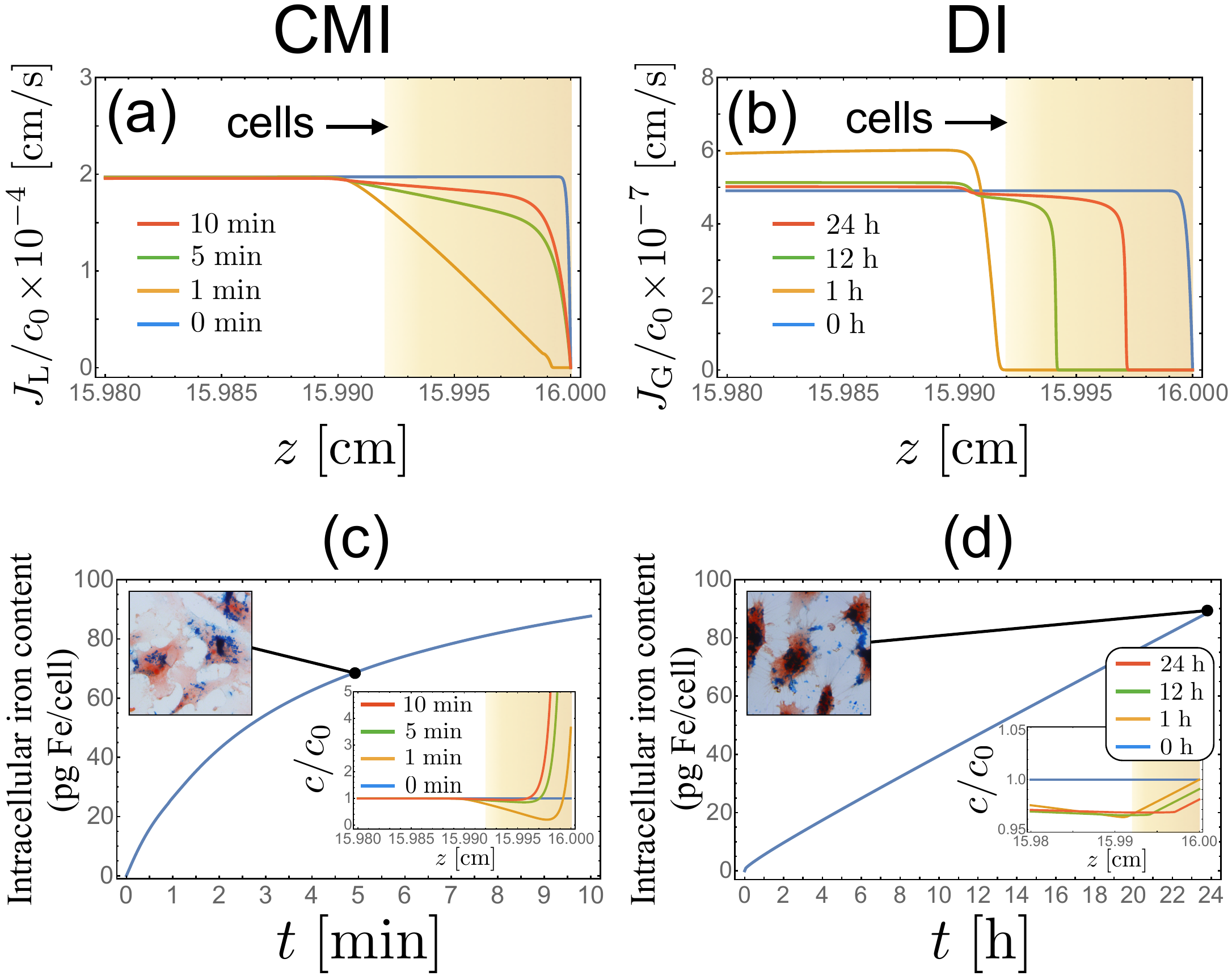}
\caption{\small Cellular internalization kinetics of NP-DMSA mediated by centrifugation (left column) and direct incubation (right column). (a) and (b) IONP normalized flux densities at the {\blue bottom end} where the cells are pelleted (shaded regions). (c) and (d) Intracellular iron content of NP-DMSA. Insets represent the spatial profiles of the normalized IONPs concentrations and optical microscopy images of U373 cells after 5 minutes (CMI) and 24 hours (DI). All rate constants used in the numerical simulations where identical in CMI and DI (see Table  {\blue \ref{tab:ParametersDMSA}}).}
\label{fig:CellsLammGrav}
\end{figure}

\begin{figure} [t!]
\centering
\includegraphics[width=0.9\linewidth]{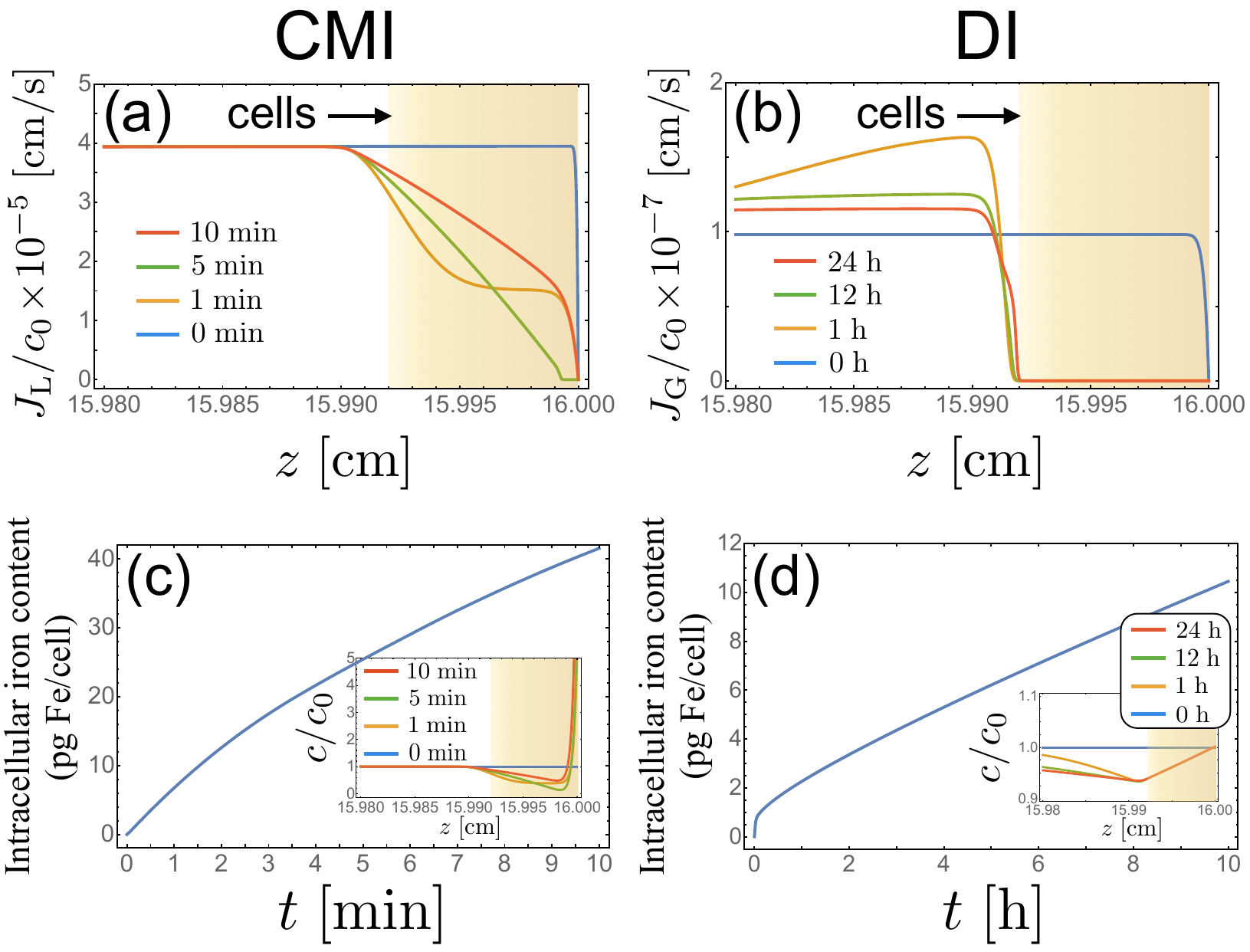}
\caption{\small Cellular internalization kinetics of NP-D mediated by centrifugation (left column) and direct incubation (right column). (a) and (b) IONP normalized flux densities at the {\blue bottom end} where the cells are pelleted (shaded regions). (c) and (d) Intracellular iron content of NP-D. Insets represent the spatial profiles of the normalized IONPs concentrations. All rate constants used in the numerical simulations where identical in CMI and DI (see Table \ref{tab:ParametersDextran}).}
\label{fig:CellsLammGrav_Dextran}
\end{figure}

\begin{figure} [t!]
\centering
\includegraphics[width=0.9\linewidth]{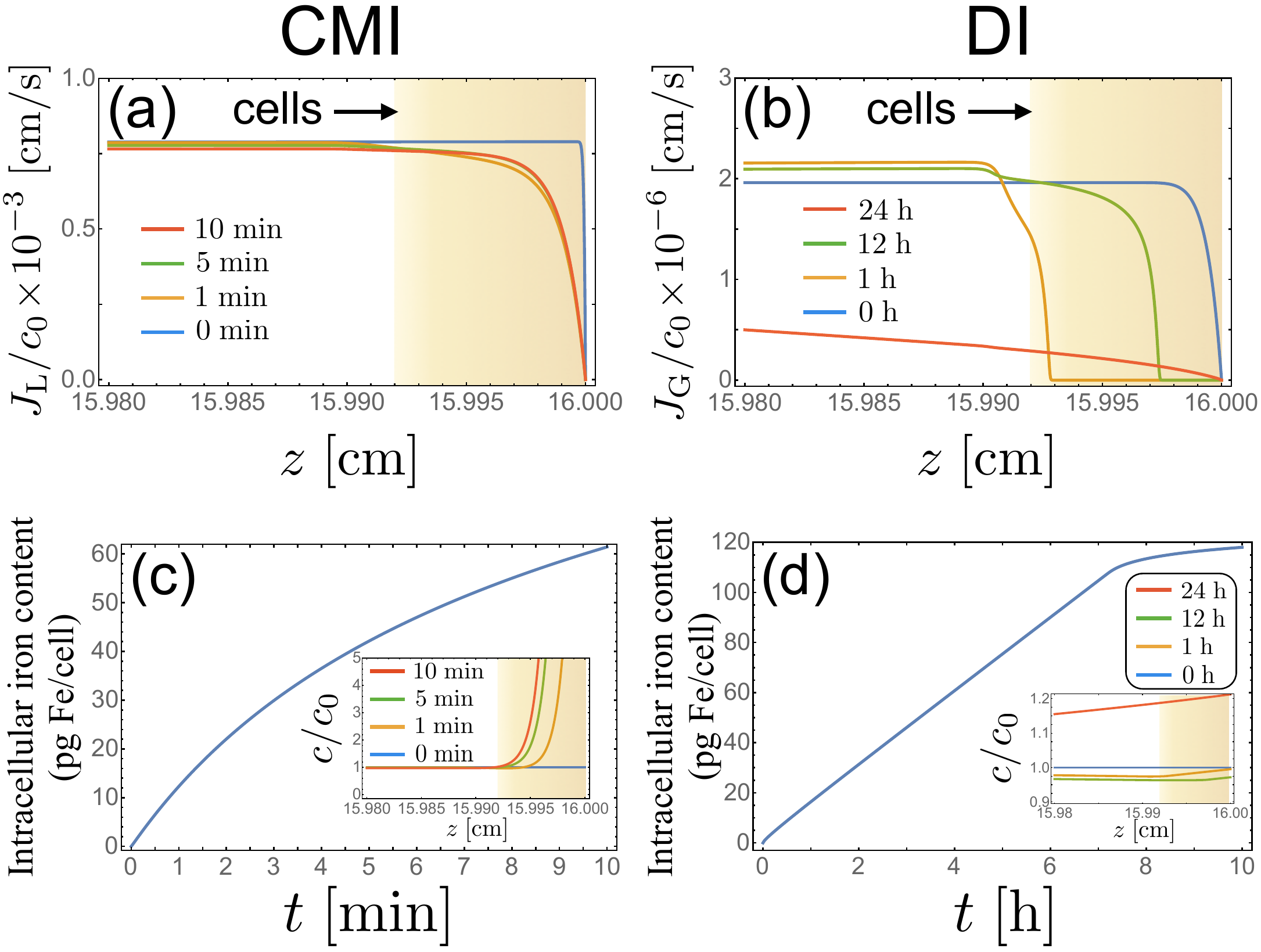}
\caption{\small Cellular internalization kinetics of NP-APS mediated by centrifugation (left column) and direct incubation (right column). (a) and (b) IONP normalized flux densities at the {\blue bottom end} where the cells are pelleted (shaded regions). (c) and (d) Intracellular iron content of NP-APS. Insets represent the spatial profiles of the normalized IONPs concentrations. All rate constants used in the numerical simulations where identical in CMI and DI (see Table \ref{tab:ParametersAPS}).}
\label{fig:CellsLammGrav_APS}
\end{figure}

Figures~\ref{fig:CellsLammGrav}, \ref{fig:CellsLammGrav_Dextran} and \ref{fig:CellsLammGrav_APS} depict the {\em in silico} kinetics of NP-DMSA, NP-D and NP-APS, respectively, both under centrifugation and gravity. They show the fluxes and masses of IONPs internalized when employing the CMI and DI methods. The role of each specific coating is addressed by adjusting both the {\blue desorption and internalization rate constants}. Notice in particular that the incoming flux density values in the outer vicinity of the cells differ by almost three orders of magnitude depending on the used configuration {\blue for NP-DMSA (Fig. \ref{fig:CellsLammGrav}) and NP-APS (Fig. \ref{fig:CellsLammGrav_APS}) coatings. Employing NP-D (Fig. \ref{fig:CellsLammGrav_Dextran}) gives rise to a flux density of two orders of magnitude higher with CMI than with DI. NP-D has more stable colloidal properties than DMSA and APS. Therefore, the use of small molecule coatings (DMSA and APS) will produce higher differences between CMI and DI when compared with macromolecules (D) due to their distinct colloidal properties, as DMSA and APS have higher hydrodynamic diameters than the dextran derivative D when employing the synthetic cell culture medium DMEM with $10\%$ of FBS}. This constitutes an important prediction of our model as it evidences the relevance played by the IONP flux density in the two configurations and dictates a first key factor to be taken into account when designing protocols to internalize in a scalable way IONPs into cells. 
\par

Figure~\ref{fig:CellsLammGrav} shows the fluxes and masses of NP-DMSA internalized when employing the CMI and DI methods. For the binding and rate constants estimated for NP-DMSA (assumed to be equal in CMI and DI), this translates into a dramatic difference between the values of the internalized IONPs masses by the cells, as evidenced in Figures~\ref{fig:CellsLammGrav}c and d. For instance, to achieve the mass uptaken by the cells during 5 minutes via the use of the CMI method it would be required to wait over 20 hours by means of the DI method. {\blue This difference is less marked for the case of NP-D (see Fig.~\ref{fig:CellsLammGrav_Dextran}) or NP-APS (see Fig.~\ref{fig:CellsLammGrav_APS}) due to their steric properties, but still large enough when comparing both internalization methods. The uptake process will be faster using CMI than DI methods in all these three coated IONPs and cannot be explained only by resorting to an interpretation based exclusively on the local values of the IONPs concentration; it requires to consider the role played by the IONPs density fluxes.} 
\par

\par

\begin{table}[H]
\caption{Kinetic parameters of NP-D}
\label{tab:ParametersDextran}
\vspace*{1mm}
\begin{tabular}{|c|c|c|}
\hline
 Symbol & Description & Range of values  \\
\hline
$k_\textrm{d}$ & Desorption rate constant & $0.32\times10^{-2}$ s$^{-1}$  \\
$k_\textrm{i}$ & Internalization rate constant &  $0.27\times10^{-1}$ s$^{-1}$  \\
$\sigma_\textrm{b}$ & Binding constant & $7.0\times10^{4}$  \\
$c_\textrm{b}^{(\textrm{max})}$ & Saturation of cell membrane concentration & 0.3 \\ 
$c_\textrm{i}^{(\textrm{max})}$ & Saturation of cell uptake concentration & 7.0 \\
\hline
\end{tabular}
\end{table}
\par

\begin{table}[H]
\caption{Kinetic parameters of NP-APS}
\label{tab:ParametersAPS}
\vspace*{1mm}
\begin{tabular}{|c|c|c|}
\hline
 Symbol & Description & Range of values  \\
\hline
$k_\textrm{d}$ & Desorption rate constant & $0.34\times10^{-2}$ s$^{-1}$  \\
$k_\textrm{i}$ & Internalization rate constant &  $0.43\times10^{-1}$ s$^{-1}$  \\
$\sigma_\textrm{b}$ & Binding constant & $7.0\times10^{4}$  \\
$c_\textrm{b}^{(\textrm{max})}$ & Saturation of cell membrane concentration & 0.3 \\ 
$c_\textrm{i}^{(\textrm{max})}$ & Saturation of cell uptake concentration & 7.0 \\
\hline
\end{tabular}
\end{table}
\par

{\blue Our model equations (\ref{eq:LammCells}) and (\ref{eq:GravCells}) predict that, by independently increasing any of the parameters $\sigma_\textrm{b}$, $k_\textrm{i}$, $c_\textrm{b}^\textrm{(max)}$ and $c_\textrm{i}^\textrm{(max)}$ or by decreasing $k_\textrm{d}$, a larger internalization mass occurs, displaying a sigmoidal-like growth profile regardless of the configuration used.} In the numerical simulations of the sets (\ref{eq:LammCells3}) and (\ref{eq:GravCells3}) we have employed the same values (see Tables \ref{table_Parameters}-\ref{tab:ParametersAPS}) for all the intervening parameters, either under centrifugation or under gravity, to provide quantitative support of the considerable differences observed in the IONPs internalization depending on the used configuration. The role of each specific coating is addressed by adjusting both the rate and binding constants. {\blue It is not discarded that these constants may exhibit slightly different values depending on the type of IONPs employed as well as on the used configuration for the container.} 
\par

\section{Conclusions}

In this paper, we have addressed a relevant mathematical problem describing the spatial-temporal dynamics of iron oxide NPs and {\blue their interaction with cells, either under moderate centrifugation or under gravity-mediated sedimentation. We put forward increasingly-complex mathematical models and solved their equations numerically using a finite element method.  Our numerical simulations showed the relevance of flow density as a key element in the interaction of NPs with the medium as well as for the processes involved in the interaction of NPs with {\em in vitro} tumor (glioma) cells. Namerly, adsorption, desorption and internalization. In addtion, our theoretical framework was accompanied by experimental results. A comparison between our experimental and numerical results was made for the average concentration of all the IONPs studied. Remarkably,  very good agreement between the numerical and experimental results was obtained}.
\par

We observed that changes in colloidal properties of the IONPs coated with small molecules (APS and DMSA) affect the sedimentation and/or diffusion coefficients, which increase, in a very short time, the concentration of IONPs in contact with the cellular membrane. In case of dextran derivatives (D and CMD), the IONPs present a good steric stabilization in the cellular media producing a lower concentration in contact with the cellular membrane. Therefore, {\blue labelling of tumor cells with the used IONPs was higher when small molecules  having colloidal properties, such as DMSA, where employed than with macromolecules, such as D and CMD. To explain these results, we extended the first transport mathematical models to elucidate the interaction of IONPs with tumor cells}, either under CMI or DI. Thus, we have quantified the role played by flux density as one of the key quantities to understand how IONPs interact with the suspension media and with cells during internalization. Using {\blue the CMI methodology a flux density of three orders of magnitude larger than the one employing the DI methodology was achieved.} 
\par

Besides, we found out that using small molecules, such as DMSA, {\blue the flux density changes in a} very short time, producing a higher amount of IONPs in contact with the cellular membrane. Hence, the internalization method of choice highly depends on the flux density in combination with the type of coating.  Our approach offers a useful framework for tailoring IONPs-based biomedical applications that require fast and reliable internalization into cells. {\blue It also brings up new questions on the role played by flux density, together with coating, for controlling the uptake of IONPs by living cells within different tissues of {\em in vivo} models.} 
\par

We hope that the present work will stimulate new mathematical models to study the use of nanoparticles as a part of therapies against tumors.

\par



\section*{Acknowledgments}

This work was supported by Fondo de Investigaciones Sanitarias, FIS [PI14/00077] and the Miguel Servet Program [CP11/00147 and CPII16/00056] from Instituto de Salud Carlos III (AAS); RTC-2015-3846-1 (AAS) and RTC-2016-4990-1 (AAS) from Spanish Ministerio de Econom\'{\i}a y Competitividad (MINECO)/FEDER funds. GS gratefully acknowledges projects MAT2015-71806-R from MINECO and NANOFRONTMAG, S2013/MIT-2850 from Madrid Regional Government (GS). BCL acknowledges MINECO (FPI program fellow-ship) from Spain. GFC and JBB thank funding from projects supported by the MINECO/FEDER MTM2015-71200-R], and the James S. Mc. Donnell Foundation 21st Century Science Initiative in Mathematical and Complex Systems Approaches for Brain Cancer (USA) [Collaborative Award 220020560].

 \appendix

 \section{Mathematical Model}

\par 

We provide below a detailed derivation and analysis of the transport partial differentials equations (PDEs) put forward in this work to model the sedimentation and diffusion of IONPs both in the absence and presence of cells in the two experimental configurations that were employed: CMI (centrifugation-mediated internalization) and DI (direct incubation). 

\subsection*{Derivation of the PDE for centrifugation-mediated sedimentation of IONPs}

\begin{figure}[h!]
\centering
\includegraphics[width=0.8\linewidth]{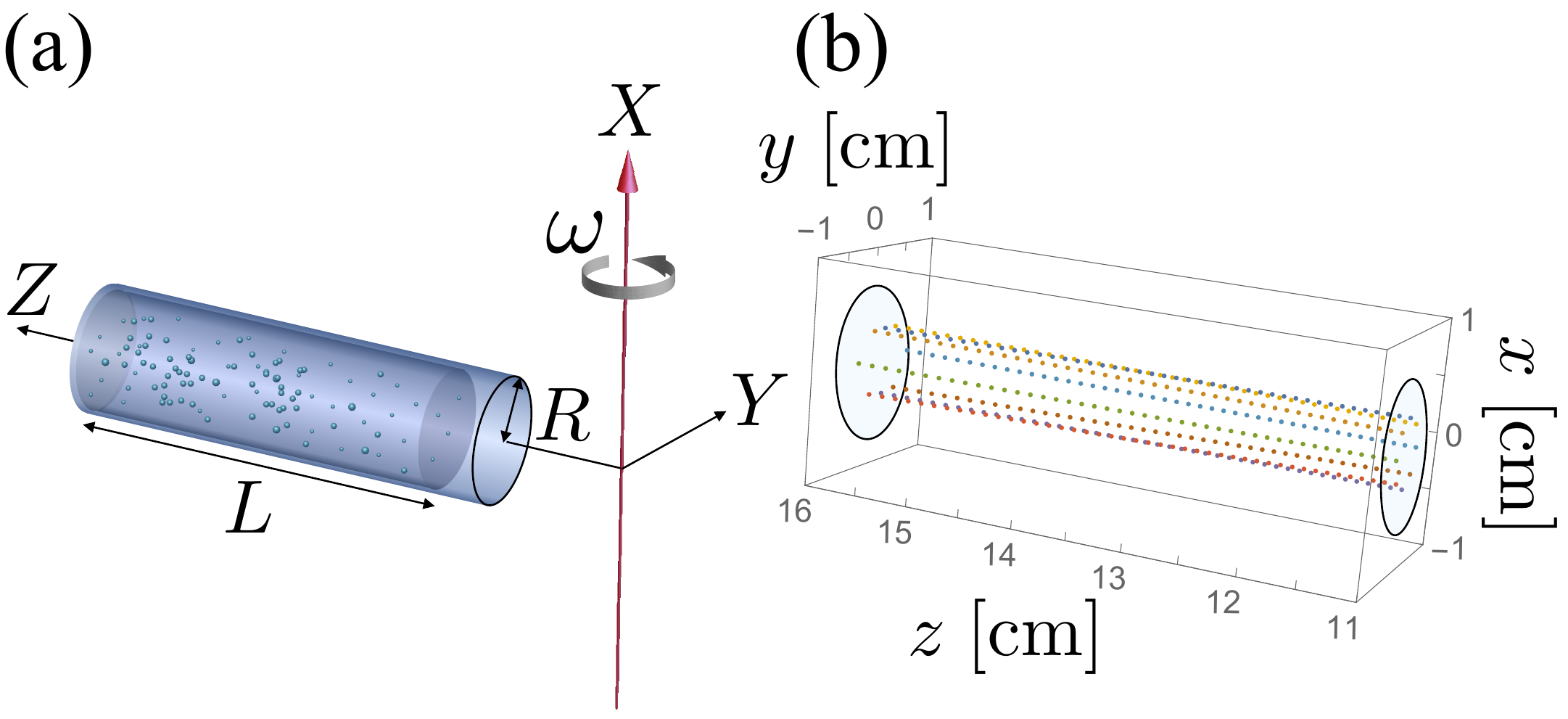}
\vspace*{-2mm}
\caption{(a) Configuration of a rotating tube containing the IONPs. (b) Trajectories of eight IONPs calculated from the numerical solution of system (\ref{eq:ODECentrifCompSimp}) with $\omega = 1500$ rpm, $\beta = 1.17\times10^{10}$ s$^{-1}$, $\sigma = 0.808$, $z_{a}=11$ cm, $z_{b}=16$ cm, and using initial conditions $r(0)=0.3$ cm, $z(0)=11$ cm, for different initial angles $\phi(0)=j\frac{\pi}{4}$, $j=0,1,\ldots,7$. The simulated time interval is $61$ hours, which is approximately the time that would be required by the IONPs, from rest, to travel from the meniscus to the the bottom end. The stochastic Brownian motion  has not been included in the trajectories for clarity.}
\label{fig:Geometry}
\end{figure}

We begin our modeling analysis by presenting a throughout derivation of the partial differential equation (\ref{eq:LammCylZ}) obeyed by the IONPs concentration $c(z,t)$ under centrifugation. We start by first defining two sets of three orthogonal unitary vectors $\left\{ {\bf u}_{x}, {\bf u}_{y}, {\bf u}_{z}\right\}$ and $\left\{ {\bf u}_{r},{\bf u}_{\phi},{\bf u}_{z}\right\}$. They correspond to Cartesian and cylindrical coordinates, respectively. Consider the rotation of a cylindrical container (a tube), of length $L=z_{b}-z_{a}$ and radius $R$, around the $X$ axis with an angular velocity vector ${\boldsymbol\omega} = \omega {\bf u}_{x}$, with $\omega$ being the constant modulus of the angular velocity. The longitudinal symmetry axis $Z$ of the tube is positioned perpendicularly to the rotation axis $X$. Figure \ref{fig:Geometry}(a) summarizes a simplified representation of the configuration of the system.
\par

Consider an IONP of mass $m_\textrm{p}$ immersed in a fluid medium and denote its position vector with respect to a rotating reference frame fixed along the tube by ${\bf r}(t) = r{\bf u}_{r} + z{\bf u}_{z}$. The mass of the fluid volume displaced by the IONP is $m_\textrm{sol}$.
\par
The IONP is subjected to a number of forces and its dynamics, in the absence of electromagnetic interactions and Brownian motion, is governed by the following vector differential equation 
\begin{eqnarray}
m_\textrm{p} \frac{d^{2}{\bf r}}{dt^{2}} \!\!\! &=&\!\!\! -\gamma_\textrm{fr} \frac{d{\bf r}}{dt} + m_\textrm{p}\,{\bf g} - m_\textrm{sol}\,{\bf g} - m_\textrm{p}\,{\boldsymbol\omega}\times\left( {\boldsymbol\omega}\times{\bf r}\right) + m_\textrm{sol}\,{\boldsymbol\omega}\times\left( {\boldsymbol\omega}\times{\bf r}\right)\nonumber\\
  \!\!\! &-&\!\!\! 2m_\textrm{p}\,{\boldsymbol\omega}\times\frac{d{\bf r}}{dt} + 2m_\textrm{sol}\,{\boldsymbol\omega}\times\frac{d{\bf r}}{dt}\, ,
\label{eq:ODECentrif}
\end{eqnarray}  
where $\gamma_\textrm{fr}$ is the friction coefficient and ${\bf g} = -g{\bf u}_{x}$ the acceleration vector due to gravity. The physical meaning of each of the right-hand-side terms of Eq. (\ref{eq:ODECentrif}) is: 
\par
\begin{itemize}
\item $-\gamma_\textrm{fr} \frac{d{\bf r}}{dt}$ is the friction force.
\item $m_\textrm{p}\,{\bf g}$ is the weight of the IONP.
\item $m_\textrm{sol}\,{\bf g}$ is the buoyancy of the IONP.
\item $- m_\textrm{p}\,{\boldsymbol\omega}\times\left( {\boldsymbol\omega}\times{\bf r}\right)$ is the centrifugal force.
\item $m_\textrm{sol}\,{\boldsymbol\omega}\times\left( {\boldsymbol\omega}\times{\bf r}\right)$ is the centrifugal buoyancy.
\item $-2m_\textrm{p}\,{\boldsymbol\omega}\times\frac{d{\bf r}}{dt}$ is the Coriolis force.
\item $2m_\textrm{sol}\,{\boldsymbol\omega}\times\frac{d{\bf r}}{dt}$ is the Coriolis buoyancy.
\end{itemize}

It is now convenient to decompose Eq. (\ref{eq:ODECentrif}) into the scalar components corresponding to the unitary vectors ${\bf u}_{r}$, ${\bf u}_{\phi}$ and ${\bf u}_{z}$, which remain fixed to the tube. Since ${\bf u}_{r} = \cos\phi\,{\bf u}_{x} + \sin\phi\, {\bf u}_{y}$, ${\bf u}_{\phi} = -\sin\phi\, {\bf u}_{x} + \cos\phi\, {\bf u}_{y}$, ${\bf r} = r{\bf u}_{r} + z{\bf u}_{z}$, ${\boldsymbol\omega}=\omega\left(\cos\phi\,{\bf u}_{r} - \sin\phi\,{\bf u}_{\phi}\right)$ and ${\bf g} = -g\left(\cos\phi\,{\bf u}_{r} - \sin\phi\,{\bf u}_{\phi}\right)$,  we arrive at the following set of scalar ordinary differential equations for the radial, angular and axial components, respectively,
\begin{subequations}\label{eq:ODECentrifComp}
\begin{eqnarray}
m_\textrm{p}\left[\frac{d^{2}r}{dt^{2}} + r\left(\frac{d{\bf\phi}}{dt}\right)^{2}\right]\!\!\! &=&\!\!\! -\gamma_\textrm{fr} \frac{dr}{dt} - \left( m_\textrm{p}-m_\textrm{sol}\right)g\cos\phi \nonumber \\
\!\!\! &+&\!\!\!  \left( m_\textrm{p}-m_\textrm{sol}\right)\omega^{2}r\sin^{2}\phi\nonumber \\
\!\!\! &+&\!\!\! 2\left( m_\textrm{p}-m_\textrm{sol}\right)\omega\sin\phi\frac{dz}{dt}\, , \label{eq:ODECentrifComp1}\\
m_\textrm{p}\left(\! r\frac{d^{2}{\bf\phi}}{dt^{2}} + 2\frac{dr}{dt}\frac{d{\bf\phi}}{dt}\!\right) \!\!\! &=&\!\!\! -\gamma_\textrm{fr} r\frac{d{\bf\phi}}{dt} - \left( m_\textrm{p}-m_\textrm{sol}\right)g\sin\phi \nonumber \\
\!\!\! &-&\!\!\! \left( m_\textrm{p}-m_\textrm{sol}\right)\omega^{2}r\cos\phi\sin\phi\nonumber \\
\!\!\! &-&\!\!\! 2\left( m_\textrm{p}-m_\textrm{sol}\right)\omega\cos\phi\frac{dz}{dt}\, , \label{eq:ODECentrifComp2}\\
m_\textrm{p}\frac{d^{2}z}{dt^{2}} \!\!\! &=&\!\!\! -\gamma_\textrm{fr} \frac{dz}{dt} + \left( m_\textrm{p}-m_\textrm{sol}\right)\omega^{2}z\nonumber \\
\!\!\! &-&\!\!\! 2\left( m_\textrm{p}-m_\textrm{sol}\right)\omega\!\left(\! \sin\phi\frac{dr}{dt} - r\cos\phi\frac{d\phi}{dt}\!\right)\! . \label{eq:ODECentrifComp3}
\end{eqnarray}  
\end{subequations}
These ordinary differential equations must be supplemented with initial conditions for the coordinates, $r(t=0)=r_{0}$, $\phi(t=0)=\phi_{0}$, $z(t=0)=z_{0}$, and the velocities $\frac{dr}{dt}(t=0)=v_{r,0}$, $\frac{d\phi}{dt}(t=0)=v_{\phi,0}$, $\frac{dz}{dt}(t=0)=v_{z,0}$.
\par
The above Eqs.~(\ref{eq:ODECentrifComp}) can be rewritten in a more handy form by taking into account Stokes' law $\gamma_\textrm{fr}=3\pi\mu d_\textrm{H}$, with $\mu$ and $d_\textrm{H}$ being the viscosity coefficient and the hydrodynamic diameter, respectively. Also, $m_\textrm{p}=\frac{\pi}{6}d_\textrm{p}^{3}\rho_\textrm{p}$ and $m_\textrm{sol}=\frac{\pi}{6}d_\textrm{p}^{3}\rho_\textrm{sol}$, where $\rho_\textrm{p}$ and $\rho_\textrm{sol}$ are the mass densities of the IONPs and the solvent, and $d_\textrm{p}$ is the geometric diameter of the IONPs, which are assumed to be spherical. Hence, we have
\begin{subequations}\label{eq:ODECentrifCompSimp}
\begin{eqnarray}
\frac{d^{2}r}{dt^{2}} + r\left(\frac{d{\bf\phi}}{dt}\right)^{2}\!\!\! &=&\!\!\! -\beta \frac{dr}{dt} - \sigma g\cos\phi + \sigma\omega^{2}r\sin^{2}\phi \nonumber\\
\!\!\! &+&\!\!\! 2\sigma\omega\sin\phi\frac{dz}{dt}\, , \label{eq:ODECentrifCompSimp1}\\
\frac{d^{2}z}{dt^{2}} \!\!\! &=&\!\!\! -\beta \frac{dz}{dt} + \sigma\omega^{2}z - 2\sigma\omega\left( \sin\phi\frac{dr}{dt} - r\cos\phi\frac{d\phi}{dt}\right)\! , \label{eq:ODECentrifCompSimp2}\\
r\frac{d^{2}{\bf\phi}}{dt^{2}} + 2\frac{dr}{dt}\frac{d{\bf\phi}}{dt} \!\!\! &=&\!\!\! -\beta r\frac{d{\bf\phi}}{dt} - \sigma g\sin\phi - \sigma\omega^{2}r\cos\phi\sin\phi \nonumber\\
\!\!\! &-&\!\!\! 2\sigma\omega\cos\phi\frac{dz}{dt}\, , \label{eq:ODECentrifCompSimp3}
\end{eqnarray}  
\end{subequations}
where we have defined $\beta = \frac{\gamma_\textrm{fr}}{m_\textrm{p}}$ and $\sigma = 1 -\frac{\rho_\textrm{sol}}{\rho_\textrm{p}}$. From these two quantities, we can express the sedimentation coefficient as
\begin{eqnarray}
s = \frac{\sigma}{\beta} = \left(\rho_{\textrm{p}} -\rho_{\textrm{sol}}\right)\! \left( \frac{d_{\textrm{p}}^{3}}{18\mu d_{\textrm{H}}}\right)\! . \label{eq:SedimentationCoefficient}
\end{eqnarray}
\par
Equations (\ref{eq:ODECentrifCompSimp}) can be solved numerically by resorting to a standard Runge-Kutta method. Figure \ref{fig:Geometry}(b) depicts a set of IONP trajectories calculated from (\ref{eq:ODECentrifCompSimp}) using $\beta = 1.17\times10^{10}$ s$^{-1}$ and $\sigma = 0.808$ (i.e. a sedimentation coefficient $s=6.91\times10^{-11}$ s). 
\par
Let $v_{r}=\frac{dr}{dt}$, $v_{\phi}=r\frac{d\phi}{dt}$ and $v_{z}=\frac{dz}{dt}$ denote the radial, azimuthal and axial velocities, respectively. In order to obtain explicit formulas for them, we first point out that for the range of parameters of our IONPs, $\beta\in[10^{9},10^{13}]$ and $\sigma\in[0.8,0.85]$, implying that $s\in[10^{-13},10^{-9}]$ s. Next, we notice that the acceleration terms $\frac{d^{2}r}{dt^{2}}$, $r\frac{d^{2}{\bf\phi}}{dt^{2}}$ and $\frac{d^{2}z}{dt^{2}}$ in (\ref{eq:ODECentrifCompSimp}) vanish within a characteristic time of $10^{-10}-10^{-8}$ s due to relaxation. Therefore, the system (\ref{eq:ODECentrifCompSimp}) can be approximated by
\begin{subequations}\label{eq:ODECentrifCompSimpVEL}
\begin{eqnarray}
v_{r}\!\!\! &=&\!\!\! - sg\cos\phi + s\omega^{2}r\sin^{2}\phi + 2s\omega\sin\phi\, v_{z} , \label{eq:ODECentrifCompSimpVEL1}\\
v_{\phi} \!\!\! &=&\!\!\! - sg\sin\phi - s\omega^{2}r\cos\phi\sin\phi - 2s\omega\cos\phi\, v_{z} , \label{eq:ODECentrifCompSimpVEL2}\\
v_{z} \!\!\! &=&\!\!\! s\omega^{2}z - 2s\omega\left( \sin\phi\, v_{r} - \cos\phi\, v_{\phi}\right) . \label{eq:ODECentrifCompSimpVEL3}
\end{eqnarray}  
\end{subequations}
Multiplying (\ref{eq:ODECentrifCompSimpVEL1}) by $\sin\phi$, (\ref{eq:ODECentrifCompSimpVEL2}) by $\cos\phi$ and subtracting, we obtain from (\ref{eq:ODECentrifCompSimpVEL3}) the following expression for the IONP velocity along the $Z$ axis
\begin{eqnarray}
v_{z} = \frac{s\omega^{2}z - 2s^{2}\omega^{3}r\sin\phi}{1+4s^{2}\omega^{2}}\, , \label{eq:ODECentrifCompSimpVELZ}
\end{eqnarray}  
where we have used the above definition (\ref{eq:SedimentationCoefficient}) for $s$. Now, since $z \gtrsim \vert r\sin\phi \vert \sim$ cm and $s\omega <10^{-5}$, which holds if $s < 10^{-8}$ and $\omega < 10^{4}$ rpm, we can accurately approximate (\ref{eq:ODECentrifCompSimpVELZ}) by
\begin{eqnarray}
v_{z} \simeq s\omega^{2}z \, . \label{eq:ODECentrifCompSimpVELZFINAL}
\end{eqnarray}
This expression displays the usual dependence of the sedimentation velocity~\cite{Schuck}. The magnitude of this velocity is in the range $v_{z}\in[10^{-8},10^{-5}]$ cm/s.
Only under ultracentrifugation conditions, where $\omega > 10^{4}$ rpm, it may be necessary to fully retain (\ref{eq:ODECentrifCompSimpVELZ}).
\par

Recalling that $v_{z}=\frac{dz}{dt}$, we can easily solve the differential equation $\frac{dz}{dt}=s\omega^{2}z$, and get $z(t)=z_{0}e^{s\omega^{2}t}$. Despite being an increasingly exponential function, due to the smallness of $s\omega^{2}$, the axial position grows very slowly, as Figure \ref{fig:Geometry}(b) illustrates. 
\par
To conclude the analysis of the velocity components, from (\ref{eq:ODECentrifCompSimpVEL1}) and (\ref{eq:ODECentrifCompSimpVEL2}), together with the above estimates and the fact that $sg\in[10^{-9},10^{-6}]$ cm/s, it follows that 
\begin{eqnarray}
v_{r}\!\!\! &\simeq&\!\!\! -sg\cos\phi + s\omega^{2}r\sin^{2}\phi\, , \label{eq:ODECentrifCompSimpVEL1FINAL}\\
v_{\phi} \!\!\! &\simeq&\!\!\! -sg\sin\phi - s\omega^{2}r\cos\phi\sin\phi \, . \label{eq:ODECentrifCompSimpVEL2FINAL}
\end{eqnarray} 
These velocities are typically one to two orders of magnitude smaller than $v_{z}$. Moreover, notice that the Coriolis terms can be neglected in (\ref{eq:ODECentrifCompSimpVEL1FINAL}) and (\ref{eq:ODECentrifCompSimpVEL2FINAL}).
\par
Having obtained explicit expressions for the three velocity components, $v_{r}$, $v_{\phi}$ and $v_{z}$, we now proceed to write down the flux densities. These combine both sedimentation and diffusion. Here we assume that the IONPs within the solvent exhibit Fickian diffusion. We will denote by $c=c(r,\phi,z,t)$ the mass concentration (density) of IONPs at spatial points $(r,\phi,z)$ and time $t$. The radial, azimuthal and axial flux densities are given by
\begin{subequations}\label{eq:Js}
\begin{eqnarray}
J_{r}\!\!\! & = &\!\!\! v_{r}\,c -D\frac{\partial c}{\partial r}\, , \label{eq:Jradial}\\
J_{\phi} \!\!\! & = &\!\!\! v_{\phi}\,c -\frac{D}{r}\frac{\partial c}{\partial \phi}\, , \label{eq:Jazimuthal}\\
J_{z}\!\!\! & = &\!\!\! v_{z}\,c -D\frac{\partial c}{\partial z}\, , \label{eq:Jz}
\end{eqnarray} 
\end{subequations}
where $D$ denotes the diffusion coefficient and $v_{r}$, $v_{\phi}$ and $v_{z}$ are provided by (\ref{eq:ODECentrifCompSimpVEL1FINAL}), (\ref{eq:ODECentrifCompSimpVEL2FINAL}) and (\ref{eq:ODECentrifCompSimpVELZFINAL}), respectively. The diffusion coefficient can be estimated from the well-known Stokes-Einstein relation
\begin{eqnarray}
D = \frac{k_\textrm{B}T}{3\pi\mu d_\textrm{H}}\, , \label{eq:Stokes-Einstein}
\end{eqnarray} 
where $k_\textrm{B}$ is Boltzmann's constant and $T$ the absolute temperature. For our IONPs, all values of the diffusion constant are $D\in[10^{-9},10^{-7}]$ cm$^{2}$/s.
\par
Since the IONPs are neither created nor destroyed during centrifugation, it follows that the concentration $c$, together with the three flux density components $J_{r}$, $J_{\phi}$ and $J_{z}$, must obey the mass continuity equation in cylindrical coordinates
\begin{eqnarray}
\frac{\partial c}{\partial t} + \frac{1}{r}\frac{\partial \left( rJ_{r}\right)}{\partial r} + \frac{1}{r}\frac{\partial J_{\phi}}{\partial \phi} + \frac{\partial J_{z}}{\partial z} = 0\, . \label{eq:ContEq}
\end{eqnarray} 
If we substitute (\ref{eq:Js}) into (\ref{eq:ContEq}) and expand all the resulting terms we arrive at the full Lamm-type equation in cylindrical coordinates
\begin{eqnarray}
\frac{\partial c}{\partial t} \!\!\! & = &\!\!\! D\!\left( \frac{\partial^{2} c}{\partial r^{2}} + \frac{1}{r} \frac{\partial c}{\partial r} + \frac{1}{r^{2}} \frac{\partial^{2} c}{\partial \phi^{2}} +\frac{\partial^{2} c}{\partial z^{2}} \right) + sg\!\left( \frac{2\cos\phi}{r} c + \cos\phi\frac{\partial c}{\partial r} + \frac{\sin\phi}{r}\frac{\partial c}{\partial\phi}\right)\nonumber\\
\!\!\! & - &\!\!\! s\omega^{2}\!\left( r\sin^{2}\phi \frac{\partial c}{\partial r} -\frac{\sin(2\phi)}{2}\frac{\partial c}{\partial\phi}+ 4\sin^{2}\phi\,c + z\frac{\partial c}{\partial z} \right)\! . \label{eq:LammCyl}
\end{eqnarray} 
This partial differential equation is supplemented with an initial condition $c(r,\phi,z,0) = c_{0}(r,\phi,z)$ for the concentration and the boundary conditions 
\begin{eqnarray}
J_{r}(R,\phi,z,t)=0\, , \quad J_{z}(r,\phi,z_{a},t)=0 \, , \quad J_{z}(r,\phi,z_{b},t)=0\, ,
\label{eq:BCLammCyl}
\end{eqnarray} 
together with the continuity of $J_{r}$ at $r=0$ and of $J_{\phi}$ for all $\phi$. The conditions (\ref{eq:BCLammCyl}) imply that no fluxes exist both at the lateral sides of the cylindrical container (at $r=R$), as well as at the meniscus (at $z=z_{a}$) and at the {\blue bottom end} (at $z=z_{b}$).
\par
Rather than solving the full Eq.~(\ref{eq:LammCyl}), it will suffice for our present purposes to focus our attention on an IONP concentration $c=c(z,t)$ depending solely on the axial variable $z$ and time $t$, and hence, irrespective of the radial and azimuthal variables $r$ and $\phi$. To derive the governing equation for $c=c(z,t)$, we define
\begin{eqnarray}
c(z,t) = \int_{0}^{R}\!\int_{0}^{2\pi}c(r,\phi,z,t)r\, dr\, d\phi\, ,
\label{eq:czt}
\end{eqnarray}
and integrate the continuity equation (\ref{eq:ContEq}) over $r$ and $\phi$. Notice now that the dimensions of $c(z,t)$ are mass per unit length. By imposing the above boundary conditions $J_{r}(R,\phi,z,t)=0$ together with the continuity of $J_{r}$ at $r=0$ and of $J_{\phi}$ for all $\phi$, we arrive at the partial differential equation (\ref{eq:LammCylZ}) which can be cast in the equivalent form
\begin{eqnarray}
\frac{\partial c}{\partial t} = \frac{\partial}{\partial z}\!\left( D\frac{\partial c}{\partial z} -s\omega^{2}zc\right)\! ,
\label{eq:LammCylZbis}
\end{eqnarray}  
where $z\in[z_{a},z_{b}]$ and $t>0$, with $c(z,t)$ satisfying an initial condition $c(z,0) = c_{0}(z)$ and the boundary conditions 
\begin{eqnarray}
J_{z}(z_{a},t)=0 \, , \quad J_{z}(z_{b},t)=0\, .
\label{eq:BCLammCylZ}
\end{eqnarray} 
\par

One particular simple case that is amenable for analytical treatment is that of a time-independent (stationary) solution. Due to the boundary conditions~(\ref{eq:BCLammCylZ}), Eq.~(\ref{eq:LammCylZbis}) reduces to an ordinary differential equation
\begin{equation}
D\frac{\partial c}{\partial z} -s\omega^{2}zc = 0 ,
\label{eq:LammCylZStationary}
\end{equation}  
whose exact solution is given by
\begin{equation}
c(z) = \mathcal{C}_{0}\exp\!\left( \frac{s\omega^{2}z^{2}}{2D}\right)\! ,
\label{eq:LammCylZStationary2}
\end{equation}
where $\mathcal{C}_{0}$ is an integration constant that can be expressed in terms of the total mass of IONPs in the medium, $M = \int_{z_{a}}^{z_{b}} c(z) dz$. Solution~(\ref{eq:LammCylZStationary2}) reflects the fact that a strong exponential localization at the {\blue bottom end} is to be expected for the concentration of the IONPs under centrifugation.


\subsection*{Generalized PDE for centrifugation-mediated sedimentation of IONPs}

Since the experimental tube carrying the medium does not strictly have a cylindrical shape but rather consists of a cylinder plus a conical end, it may be questioned whether Eq.~(\ref{eq:LammCylZbis}) remains valid in such a setting. The configuration of the tube employed (Falcon$^\textrm{TM}$ 15 ml) is still of axial symmetry around the $Z$ axis, with a radius $R=R(z)$ that is not constant at all cross-sections as $z$ varies. Hence, we will extend (\ref{eq:czt}) and define instead
\begin{eqnarray}
c(z,t) = \int_{0}^{R(z)}\!\!\int_{0}^{2\pi}c(r,\phi,z,t)r\, dr\, d\phi\, ,
\label{eq:cztMod}
\end{eqnarray}
where we now make explicit the dependence on $z$ of the tube radius $R(z)$. 
\par

\begin{figure}[h!]
\centering
\includegraphics[width=0.5\linewidth]{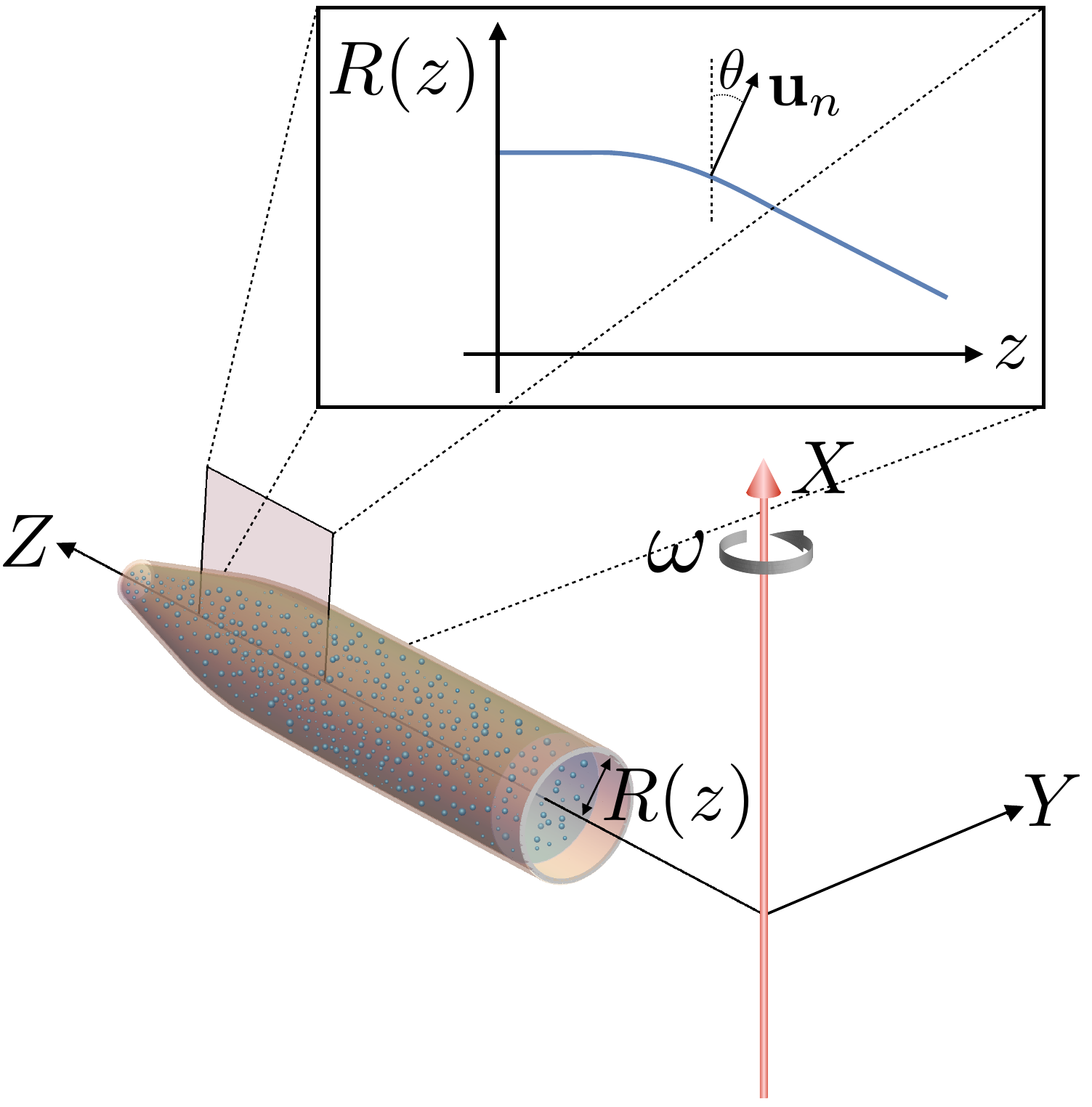}
\caption{Configuration of the container used in our experiments, where the radius $R(z)$ of the cross section varies with the distance to the centrifugation axis $X$. }
\label{fig:GeometryFalcon}
\end{figure}
\par

We consider again the continuity equation~(\ref{eq:ContEq}), as it is fully valid for a general 3D geometry, and integrate it over the coordinates $r$ and $\phi$. We have the following results for each of the intervening terms:
\begin{equation}
\int_{0}^{R(z)}\!\!\int_{0}^{2\pi}\frac{\partial c(r,\phi,z,t)}{\partial t}\, r\, dr\, d\phi = \frac{\partial c(z,t)}{\partial t},
\label{eq:cztModtime}
\end{equation}
due to the fact that the limits of integration do not depend on time. Also,
\begin{equation}
\int_{0}^{R(z)}\!\!\int_{0}^{2\pi}\left[ \frac{1}{r}\frac{\partial J_{\phi}}{\partial \phi}\right] r\, dr\, d\phi = 
\int_{0}^{R(z)} J_{\phi}(r,\phi,z,t)\big\vert_{0}^{2\pi} dr = 0\, ,\label{eq:cztModang}
\end{equation}
because of the continuity of $J_{\phi}(r,\phi,z,t)$ for all $\phi\in[0,2\pi]$.
\par
Now, to deal with the last two terms in Eq.~(\ref{eq:ContEq}), we impose the boundary condition that the normal component of the flux density vector ${\bf J} = J_{r}{\bf u}_{r} + J_{\phi}{\bf u}_{\phi} + J_{z}{\bf u}_{z}$ should be zero at $r = R(z)$. The unit outward normal vector to the surface of the tube is ${\bf u}_{n} = \cos\theta(z)\, {\bf u}_{r} + \sin\theta(z)\, {\bf u}_{z}$, where $\theta(z)$ is the subtended angle of ${\bf u}_{n}$ with the radial axis and depends on $z$ (see Figure~\ref{fig:GeometryFalcon}). Hence, from ${\bf J}\cdot{\bf u}_{n} =0$ at $r = R(z)$, we get
\begin{equation}
J_{r}(R(z),\phi,z,t) = -\tan\theta(z)\, J_{z}(R(z),\phi,z,t)\, . \label{eq:cztModJradJz}
\end{equation}
The integral for the radial component yields
\begin{eqnarray}
\int_{0}^{R(z)}\!\!\int_{0}^{2\pi}\!\left[ \frac{1}{r}\frac{\partial \left( rJ_{r}\right)}{\partial r}\right]\! r\, dr\, d\phi \!\! &= & \!\! 
\int_{0}^{2\pi} rJ_{r}(r,\phi,z,t)\big\vert_{0}^{R(z)} d\phi \nonumber\\
\!\! &=&\!\! -\tan\theta(z)\, R(z)\! \int_{0}^{2\pi} \!J_{z}(R(z),\phi,z,t)\,d\phi \, ,\nonumber\\
\label{eq:cztModrad}
\end{eqnarray}
where in the last step we have made use of (\ref{eq:cztModJradJz}).
\par
To evaluate $\int_{0}^{R(z)}\!\!\int_{0}^{2\pi}\frac{\partial J_{z}(r,\phi,z,t)}{\partial z} r\, dr\, d\phi$, we define 
\begin{eqnarray}
J_{z}(z,t) = \int_{0}^{R(z)}\!\!\int_{0}^{2\pi} J_{z}(r,\phi,z,t) r\, dr\, d\phi\, ,\label{eq:cztModJzreduced}
\end{eqnarray}
and consider
\begin{eqnarray}
\frac{\partial J_{z}(z,t)}{\partial z} \!\! &=&\!\! \frac{\partial}{\partial z}\!\int_{0}^{R(z)}\!\!\int_{0}^{2\pi} J_{z}(r,\phi,z,t) r\, dr\, d\phi =\! \int_{0}^{R(z)}\!\!\int_{0}^{2\pi}\frac{\partial J_{z}(r,\phi,z,t)}{\partial z} r\, dr\, d\phi  \nonumber\\
\!\! &+&\!\!  R(z)\frac{dR}{dz}\int_{0}^{2\pi} J_{z}(R(z),\phi,z,t)\,d\phi\, ,\label{eq:cztModJzreducedINT}
\end{eqnarray}
where in the last step we have resorted to Leibniz's rule for differentiation under the integral sign.
\par
Therefore, from the continuity equation~(\ref{eq:ContEq}), and combining (\ref{eq:cztModtime}), (\ref{eq:cztModang}), (\ref{eq:cztModrad}) and (\ref{eq:cztModJzreducedINT}), we arrive at
\begin{equation}
\frac{\partial c(z,t)}{\partial t} + \frac{\partial J_{z}(z,t)}{\partial z} - \left( \tan\theta(z) + \frac{dR}{dz}\right)\!R(z)\int_{0}^{2\pi} J_{z}(r,\phi,z,t) r\, dr\, d\phi = 0\, .\label{eq:cztMod2}
\end{equation}
Now, notice from the inset in Figure~\ref{fig:GeometryFalcon} that $\tan\theta(z) = -\frac{dR}{dz}$. Hence, (\ref{eq:cztMod2}) exactly collapses to the following continuity equation
\begin{eqnarray}
\frac{\partial c(z,t)}{\partial t} + \frac{\partial J_{z}(z,t)}{\partial z} = 0\, . \label{eq:cztMod3}
\end{eqnarray}
To bring out the concentration $c(z,t)$ in Eq.~(\ref{eq:cztMod3}), we use the definition (\ref{eq:cztModJzreduced}) together with the fact, see Eq.~(\ref{eq:Jz}), that
\begin{eqnarray}
J_{z}(z,t) \!\! &=&\!\! \int_{0}^{R(z)}\!\!\int_{0}^{2\pi} \left( s\omega^{2} zc(r,\phi,z,t)- D \frac{\partial c(r,\phi,z,t)}{\partial z}\right) r\, dr\, d\phi \nonumber\\
\!\! &=&\!\! s\omega^{2} z\,c(z,t)- D\!\int_{0}^{R(z)}\!\!\int_{0}^{2\pi} \frac{\partial c(r,\phi,z,t)}{\partial z}\, r\, dr\, d\phi
\, ,\label{eq:cztModJzreduced2}
\end{eqnarray}
where in the last step we have employed (\ref{eq:czt}). To evaluate the last integral in (\ref{eq:cztModJzreduced2}) we consider
\begin{eqnarray}
\frac{\partial c(z,t)}{\partial z} \!\! &=&\!\! \frac{\partial}{\partial z}\int_{0}^{R(z)}\!\!\int_{0}^{2\pi} c(r,\phi,z,t) r\, dr\, d\phi =\int_{0}^{R(z)}\!\!\int_{0}^{2\pi}\frac{\partial c(r,\phi,z,t)}{\partial z}\, r\, dr\, d\phi \nonumber\\
\!\! &+&\!\!   R(z)\frac{dR}{dz}\int_{0}^{2\pi} c(R(z),\phi,z,t)\,d\phi\, ,\label{eq:cztMod4}
\end{eqnarray}
where we have applied once more Leibniz's rule.
\par
Plugging expressions (\ref{eq:cztModJzreduced2}) and (\ref{eq:cztMod4}) into the continuity equation (\ref{eq:cztMod3}), we obtain
\begin{equation}
\frac{\partial c(z,t)}{\partial t} =  \frac{\partial}{\partial z}\left( D\frac{\partial c(z,t)}{\partial z} - s\omega^{2} zc(z,t) -DR\frac{dR}{dz}\int_{0}^{2\pi} c(R,\phi,z,t)\,d\phi\right)\! . \label{eq:cztMod5}
\end{equation}
This partial differential equation, supplemented with the boundary conditions $J_{z}(z_{a},t)=0$ and $J_{z}(z_{b},t)=0$, provides the generalization for a medium container possessing axial symmetry about the $Z$ axis. It is apparent that the only difference between the continuity equation (\ref{eq:cztMod5}) and (\ref{eq:LammCylZbis}) is the presence of a new term, involving the diffusion coefficient, that accounts for the variation of the radius $R(z)$ with the axial distance $z$. 
\par

Up to this point, and within our model framework, all calculations have been exact. To assess the relevance of the correction term 
$$DR(z)\frac{dR}{dz}\int_{0}^{2\pi} c(R(z),\phi,z,t)\,d\phi$$
 in (\ref{eq:cztMod5}), we use (\ref{eq:cztMod}) and approximate the integral over the radial variable by means of the trapezoidal rule. Although this is a somewhat rough approximation, it nevertheless furnishes a good estimation of the order of magnitude of this term. We have
\begin{eqnarray}
c(z,t) = \int_{0}^{R(z)}\!\!\int_{0}^{2\pi}\!c(r,\phi,z,t)r\, dr\, d\phi \simeq \frac{R^{2}(z)}{2}\!\int_{0}^{2\pi}\! c(R(z),\phi,z,t)\,d\phi ,
\label{eq:cztMod6}
\end{eqnarray}
and thus $\int_{0}^{2\pi} c(R(z),\phi,z,t)\,d\phi \simeq \frac{2c(z,t)}{R^{2}(z)}$. Hence, Eq.~(\ref{eq:cztMod5}) is approximated by 
\begin{eqnarray}
\frac{\partial c(z,t)}{\partial t} =  \frac{\partial}{\partial z}\left( D\frac{\partial c(z,t)}{\partial z} - s\omega^{2} z\,c(z,t) -\frac{2D}{R(z)}\frac{dR}{dz}c(z,t)\right) . \label{eq:cztMod7}
\end{eqnarray}

In our experimental set up the centrifuge tube (model Falcon$^\textrm{TM}$ 15 ml) has a radius $R(z)$ that changes from $R_{a}=6.9$ mm to $R_{b}=1.8$ mm over the conical cross-section, which has a length of 24 mm (the total length of the tube is 120 mm). Moreover, since $\omega = 1500$ rpm  and for the typical $s/D$ ratios of our IONPs, it follows that 
\begin{eqnarray}
s\omega^{2} z \gg \Bigg\vert\frac{2D}{R(z)}\frac{dR}{dz}\Bigg\vert \, , \label{eq:cztMod8}
\end{eqnarray}
because the difference between both terms is within one to two orders of magnitude.
\par

Therefore, we conclude that Eq.~(\ref{eq:LammCylZbis}) provides an accurate model to describe the diffusion and sedimentation of IONPs in the axially-symmetric centrifuge tube employed in our experiments. Notice also that the action of the gravitational field, even though it has been included in our previous analysis, via Eq. (\ref{eq:ODECentrif}),  is absent in our final Eq.~(\ref{eq:LammCylZbis}). This is a consequence of the form of the axial velocity $v_{z}$, as can be observed in~(\ref{eq:ODECentrifCompSimpVELZ}) or in~(\ref{eq:ODECentrifCompSimpVELZFINAL}). As it will be described in the next subsection, an analogous spatio-temporal equation can be derived when the gravity is the only driving mechanism of sedimentation.
\par


\subsection*{Derivation of the PDE for gravitation-mediated sedimentation of IONPs}

\begin{figure}[h!]
\centering
\includegraphics[width=0.8\linewidth]{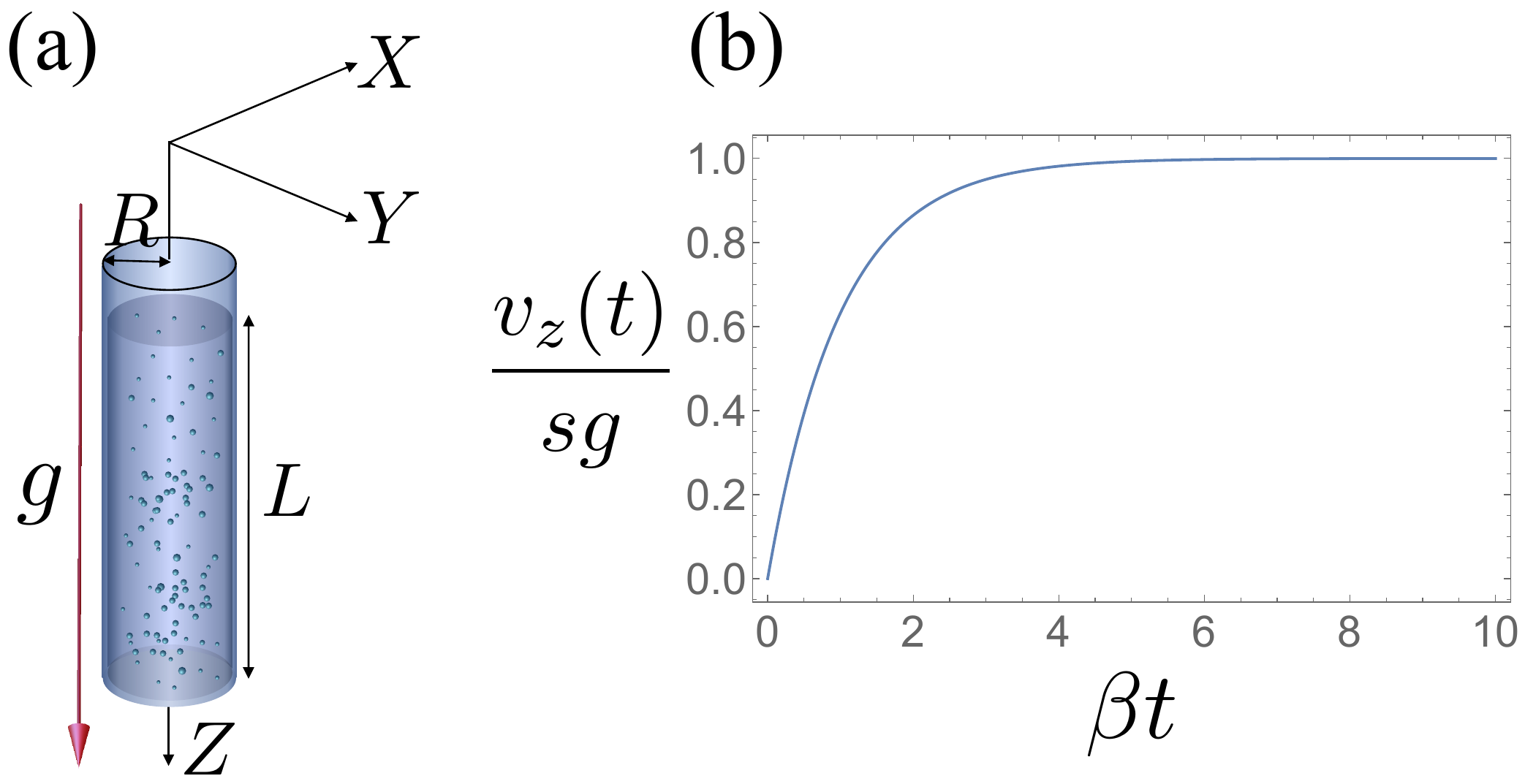}
\caption{\small (a) Configuration of the tube containing the IONPs under gravity. (b) Normalized sedimentation velocity component $v_{z}$ along the axial $Z$ direction obtained from (\ref{eq:ExactGravCompZ}). This sedimentation velocity tends to a constant value $sg$ for $t> \beta^{-1}$.}
\label{fig:GeometryGrav}
\end{figure}

In the sole presence of the gravitational force, sedimentation can also take place although at a significantly slower pace when compared with the application of centrifugation. In our experimental setting, the container (assumed henceforth to be cylindrical), comprising the solvent and the IONPs, is now placed with its longitudinal symmetry axis along the gravity field (i.e. the $Z$ axis), as shown in Figure \ref{fig:GeometryGrav}(a).
\par

We may resort to the same theoretical framework employed in previous subsections. Again, consider an IONP of mass $m_\textrm{p}$ immersed in a fluid medium and denote its position vector with respect to a reference frame fixed along the tube by ${\bf r}(t) = r{\bf u}_{r} + z{\bf u}_{z}$. The mass of the fluid volume displaced by the IONP is $m_\textrm{sol}$.
\par

The IONP is subjected to a number of forces and its dynamics, in the absence of electromagnetic interactions and Brownian motion, is governed by the following vector differential equation 
\begin{eqnarray}
m_\textrm{p} \frac{d^{2}{\bf r}}{dt^{2}} \!\!\! &=&\!\!\! -\gamma_\textrm{fr} \frac{d{\bf r}}{dt} + m_\textrm{p}\,{\bf g} - m_\textrm{sol}\,{\bf g} \, ,
\label{eq:ODEGravit}
\end{eqnarray}
where $\gamma_\textrm{fr}$ is the friction coefficient and ${\bf g} = g{\bf u}_{z}$ the acceleration vector due to gravity. The physical meaning of each of the right-hand-side terms of Eq. (\ref{eq:ODEGravit}) is:   
\begin{itemize}
\item $-\gamma_\textrm{fr} \frac{d{\bf r}}{dt}$ is the friction force.
\item $m_\textrm{p}\,{\bf g}$ is the weight of the IONP.
\item $-m_\textrm{sol}\,{\bf g}$ is the buoyancy of the IONP.
\end{itemize}
\par
We now decompose Eq. (\ref{eq:ODEGravit}) into the scalar components corresponding to the unitary vectors ${\bf u}_{r}$, ${\bf u}_{\phi}$ and ${\bf u}_{z}$, which remain fixed to the tube. Since ${\bf u}_{r} = \cos\phi\,{\bf u}_{x} + \sin\phi\, {\bf u}_{y}$ and ${\bf u}_{\phi} = -\sin\phi\, {\bf u}_{x} + \cos\phi\, {\bf u}_{y}$,  we arrive at the following set of scalar ordinary differential equations for the radial, angular and axial components, respectively,
\begin{subequations}\label{eq:ODEGravitComp}
\begin{eqnarray}
m_\textrm{p}\left[\frac{d^{2}r}{dt^{2}} + r\left(\frac{d{\bf\phi}}{dt}\right)^{2}\right]\!\!\! &=&\!\!\! -\gamma_\textrm{fr} \frac{dr}{dt} \, , \label{eq:ODEGravitComp1}\\
m_\textrm{p}\left( r\frac{d^{2}{\bf\phi}}{dt^{2}} + 2\frac{dr}{dt}\frac{d{\bf\phi}}{dt}\right) \!\!\! &=&\!\!\! -\gamma_\textrm{fr} r\frac{d{\bf\phi}}{dt} \, , \label{eq:ODEGravitComp2}\\
m_\textrm{p}\frac{d^{2}z}{dt^{2}} \!\!\! &=&\!\!\! -\gamma_\textrm{fr} \frac{dz}{dt} + \left( m_\textrm{p}-m_\textrm{sol}\right)g . \label{eq:ODEGravitComp3}
\end{eqnarray}  
\end{subequations}
These ordinary differential equations must be supplemented with initial conditions for the coordinates, $r(t=0)=r_{0}$, $\phi(t=0)=\phi_{0}$, $z(t=0)=z_{0}$, and the velocities $\frac{dr}{dt}(t=0)=v_{r,0}$, $\frac{d\phi}{dt}(t=0)=v_{\phi,0}$, $\frac{dz}{dt}(t=0)=v_{z,0}$. In contrast with Eqs.(\ref{eq:ODECentrifComp}), in (\ref{eq:ODEGravitComp}) the axial component is uncoupled to the radial and azimuthal components.
\par
The above Eqs.~(\ref{eq:ODEGravitComp}) can be rewritten in a more convenient form by taking into account that $\gamma_\textrm{fr}=3\pi\mu d_\textrm{H}$, with $\mu$ and $d_\textrm{H}$ being the viscosity coefficient and the hydrodynamic diameter, respectively. Also, $m_\textrm{p}=\frac{\pi}{6}d_\textrm{p}^{3}\rho_\textrm{p}$ and $m_\textrm{sol}=\frac{\pi}{6}d_\textrm{p}^{3}\rho_\textrm{sol}$, where $\rho_\textrm{p}$ and $\rho_\textrm{sol}$ are the mass densities of the IONPs and the solvent, and $d_\textrm{p}$ is the geometric diameter of the IONPs, which are assumed to be spherical. Hence, we have
\begin{subequations}\label{eq:ODEGravitCompSimp}
\begin{eqnarray}
\frac{d^{2}r}{dt^{2}} + r\left(\frac{d{\bf\phi}}{dt}\right)^{2}\!\!\! &=&\!\!\! -\beta \frac{dr}{dt} \, , \label{eq:ODEGravitCompSimp1}\\
r\frac{d^{2}{\bf\phi}}{dt^{2}} + 2\frac{dr}{dt}\frac{d{\bf\phi}}{dt} \!\!\! &=&\!\!\! -\beta r\frac{d{\bf\phi}}{dt} \, , \label{eq:ODEGravitCompSimp2}\\
\frac{d^{2}z}{dt^{2}} \!\!\! &=&\!\!\! -\beta \frac{dz}{dt} + \sigma g . \label{eq:ODEGravitCompSimp3}
\end{eqnarray}  
\end{subequations}
where we have defined $\beta = \frac{\gamma_\textrm{fr}}{m_\textrm{p}}$ and $\sigma = 1 -\frac{\rho_\textrm{sol}}{\rho_\textrm{p}}$. The sedimentation coefficient, $s=\frac{\sigma}{\beta}$, is provided by expression (\ref{eq:SedimentationCoefficient}). 
\par
The exact solution to Eq.~(\ref{eq:ODEGravitCompSimp3}) is given by 
\begin{eqnarray}
z(t) = z_{0} + \frac{v_{z,0}}{\beta} - \frac{sg}{\beta} + sgt + \frac{sg}{\beta}e^{-\beta t} - \frac{v_{z,0}}{\beta} e^{-\beta t}. \label{eq:ExactGravCompZ}
\end{eqnarray} 

Just as it occurred with the centrifugation scenario, there is a very fast relaxation. Thus, for $t >\beta^{-1}$, all the velocity components are essentially constant. The radial and the azimuthal components $v_{r}=0$ and $v_{\phi}=0$, whereas the axial component is equal to (see Figure \ref{fig:GeometryGrav}.b)
\begin{eqnarray}
v_{z} =  sg\, . \label{eq:ODEGravCompSimpVELZ}
\end{eqnarray}  
When we compare the axial components of the velocity under centrifugation, given by (\ref{eq:ODECentrifCompSimpVELZFINAL}), and under gravity, given by (\ref{eq:ODEGravCompSimpVELZ}), we see that the former one is $\frac{\omega^{2}z}{g}$ times larger. For $10^{3} < \omega < 10^{4}$ rpm and $z\sim$ cm, the axial velocity under centrifugation is in the order of $10^2-10^3$ times larger than the corresponding one under the sole action of gravity.
\par
Having obtained explicit expressions for the three velocity components, $v_{r}$, $v_{\phi}$ and $v_{z}$, we now proceed to write down the flux densities. These comprise both sedimentation and diffusion. Just as we did with the case of centrifugation, we assume that the IONPs within the solvent exhibit Fickian diffusion with a diffusion coefficient $D$. We will denote by $c=c(r,\phi,z,t)$ the mass concentration (density) of IONPs at spatial points $(r,\phi,z)$ and time $t$. The radial, azimuthal and axial flux densities have identical structure to those given by Eqs.(\ref{eq:Js}). However, since $v_{r}=0$, $v_{\phi}=0$ and $v_{z} =  sg$, we obtain 
\begin{subequations}\label{eq:JsGravity}
\begin{eqnarray}
J_{r}\!\!\! & = &\!\!\! -D\frac{\partial c}{\partial r}\, , \label{eq:JradialGravity}\\
J_{\phi} \!\!\! & = &\!\!\! -\frac{D}{r}\frac{\partial c}{\partial \phi}\, , \label{eq:JazimuthalGravity}\\
J_{z}\!\!\! & = &\!\!\! sgc -D\frac{\partial c}{\partial z}\, . \label{eq:JzGravity}
\end{eqnarray} 
\end{subequations}
Since the IONPs are neither created nor destroyed during gravitation-media-\\ted sedimentation, it follows that the concentration $c$, together with the three flux density components $J_{r}$, $J_{\phi}$ and $J_{z}$, must obey the mass continuity equation in cylindrical coordinates (\ref{eq:ContEq}). If we substitute (\ref{eq:JsGravity}) into (\ref{eq:ContEq}) and expand all the resulting terms we arrive at the full diffusion-gravity equation in cylindrical coordinates
\begin{eqnarray}
\frac{\partial c}{\partial t} \!\!\! & = &\!\!\! D\!\left( \frac{\partial^{2} c}{\partial r^{2}} + \frac{1}{r} \frac{\partial c}{\partial r} + \frac{1}{r^{2}} \frac{\partial^{2} c}{\partial \phi^{2}} +\frac{\partial^{2} c}{\partial z^{2}} \right) - sg\frac{\partial c}{\partial z}. \label{eq:DiffGravityCyl}
\end{eqnarray} 
This partial differential equation is supplemented with an initial condition $c(r,\phi,z,0) = c_{0}(r,\phi,z)$ for the concentration and the boundary conditions 
\begin{eqnarray}
J_{r}(R,\phi,z,t)=0\, , \quad J_{z}(r,\phi,z_{a},t)=0 \, , \quad J_{z}(r,\phi,z_{b},t)=0\, ,
\label{eq:BCGravityCyl}
\end{eqnarray} 
together with the continuity of $J_{r}$ at $r=0$ and of $J_{\phi}$ for all $\phi$. These conditions are identical to (\ref{eq:BCLammCyl}), and imply that no fluxes exist both at the lateral sides of the cylindrical container as well as at the meniscus (at $z=z_{a}$) and at the {\blue bottom end} (at $z=z_{b}$).
\par
Rather than solving (\ref{eq:DiffGravityCyl}), it will suffice for our present purposes to restrict our attention to an IONP concentration $c=c(z,t)$ depending only on the axial variable $z$ and time $t$, and hence, irrespective of the radial and azimuthal variables $r$ and $\phi$. To derive the governing equation for $c=c(z,t)$, we resort again to (\ref{eq:czt}) and integrate the continuity equation (\ref{eq:ContEq}) over $r$ and $\phi$. By imposing the above boundary conditions $J_{r}(R,\phi,z,t)=0$ together with the continuity of $J_{r}$ at $r=0$ and of $J_{\phi}$ for all $\phi$, we finally arrive at the partial differential equation (\ref{eq:PDEGravity}).
\par

It is worth mentioning that the exact solution to (\ref{eq:PDEGravity}), satisfying the boundary conditions (\ref{eq:BC}), can be found via the standard method of separation of variables and reads as
\begin{eqnarray}
c(z,t)= A_{0}e^{2\kappa z} \!\!\!&+&\!\!\! e^{\kappa z - \kappa^{2} D t}\sum_{n=1}^{\infty} A_{n}\!\left[ \cos\left(\mu_{n} \left(z - z_{a}\right)\right) + \frac{\kappa}{\mu_{n}}\sin\left(\mu_{n} \left(z - z_{a}\right)\right) \right]\! \nonumber\\
\!\!\!&\times&\!\!\!e^{-\mu_{n}^{2}Dt},
\label{eq:Solgravity}
\end{eqnarray} 
where $\kappa = \frac{sg}{2D}$, $\mu_{n}=\frac{\pi n}{L}$, $L=z_{b}-z_{a}$ and the Fourier coefficients are
\begin{eqnarray*}
A_{0} \!\!\! &=&\!\!\! \frac{2\kappa}{e^{2\kappa z_{b}} - e^{2\kappa z_{a}}}\!\int_{z_{a}}^{z_{b}}\! c_{0}(z) dz \, , \label{eq:FourierCoef0}\\
A_{n} \!\!\! &=&\!\!\! \frac{2\mu_{n}^{2}}{\left(\mu_{n}^{2} + \kappa^{2}\right)\! L}\!\int_{z_{a}}^{z_{b}}\!\! c_{0}(z)\! \left[ \cos\left(\mu_{n} \left(z - z_{a}\right)\right) + \frac{\kappa}{\mu_{n}}\sin\left(\mu_{n} \left(z - z_{a}\right)\right) \right]\! e^{-\kappa z}dz,
\label{eq:FourierCoef}
\end{eqnarray*} 
with $n=1,2,\ldots$. If the initial condition $c_{0}(z)$ is uniform for all $z\in[z_{a},z_{b}]$, then, as anticipated, $c(z,t)$ remains constant for all $z\in[z_{a},z_{b}]$ and $t>0$. Expression~(\ref{eq:Solgravity}) contains a time-independent term (the asymptotic solution) and allows one to obtain the spatio-temporal concentration of IONPs for sufficiently long times as only the lowest-order terms in the Fourier expansion are the relevant ones.
\par

Notice that Eq.~(\ref{eq:PDEGravity}) can be extended to the situation where the container has an axially-symmetric shape. We start from
the continuity equation (\ref{eq:cztMod3}), which also holds under gravity, and define the corresponding axial flux density
\begin{eqnarray}
J_{z}(z,t) \!\! &=&\!\! \int_{0}^{R(z)}\!\!\int_{0}^{2\pi} \left( sg\,c(r,\phi,z,t)- D \frac{\partial c(r,\phi,z,t)}{\partial z}\right) r\, dr\, d\phi \nonumber\\
\!\! &=&\!\! sg\,c(z,t)- D\int_{0}^{R(z)}\!\!\int_{0}^{2\pi} \frac{\partial c(r,\phi,z,t)}{\partial z}\, r\, dr\, d\phi
 ,\label{eq:cztgravMod}
\end{eqnarray}
where we have used the fact that $J_{z}(r,\phi,z,t)=sg\,c(r,\phi,z,t)- D\, \frac{\partial c(r,\phi,z,t)}{\partial z}$. Following analogous steps as in the centrifugation scenario {\blue analysed} previously, we arrive at the partial differential equation 
\begin{eqnarray}
\frac{\partial c(z,t)}{\partial t} =  \frac{\partial}{\partial z}\!\left( D\frac{\partial c(z,t)}{\partial z} - sgc(z,t) -DR\,\frac{dR}{dz}\!\int_{0}^{2\pi}\! c(R,\phi,z,t)\,d\phi\right)\! .  \label{eq:cztgravMod2}
\end{eqnarray}
This partial differential equation, supplemented with the boundary conditions $J_{z}(z_{a},t)=0$ and $J_{z}(z_{b},t)=0$, provides the generalization for a tube possessing axial symmetry about the $Z$ axis. It is apparent that the only difference between the continuity equation (\ref{eq:cztgravMod2}) and (\ref{eq:PDEGravity}) is the presence of a new term, involving the diffusion coefficient, that accounts for the variation of the radius $R(z)$ with the axial distance $z$. 
\par
In principle, we cannot neglect the new term with respect to the gravity-mediated sedimentation term $sgc(z,t)$. However, since the IONP concentration will not display strong spatial variations (apart from those that may occur at the initial condition), we can approximate the integral over the radial variable by means of the trapezoidal rule to simplify the new term. From (\ref{eq:cztMod6}) we have $\int_{0}^{2\pi} c(R(z),\phi,z,t)\,d\phi \simeq \frac{2c(z,t)}{R^{2}(z)}$. Hence, Eq.~(\ref{eq:cztgravMod2}) can be approximated by 
\begin{eqnarray}
\frac{\partial c(z,t)}{\partial t} =  \frac{\partial}{\partial z}\left( D\,\frac{\partial c(z,t)}{\partial z} - sg\,c(z,t) -\frac{2D}{R(z)}\frac{dR}{dz}c(z,t)\right)\! . \label{eq:PDEGravityMod}
\end{eqnarray}
Numerically, in our studied scenario, the differences between the solutions of (\ref{eq:PDEGravityMod}) and (\ref{eq:PDEGravity}) were very small. Thus, in our simulations to compare with the DI sedimentation experiments, it was sufficient to use (\ref{eq:PDEGravity}).
\par


{\blue 

\subsection*{Extensions to IONPs distributed in sedimentation and diffusion}

In both Eqs.~(\ref{eq:LammCylZ}) and (\ref{eq:PDEGravity}) all IONPs in the solvent were assumed to have the same sedimentation and diffusion coefficients. This is justified if the IONPs have approximately equal sizes, densities and viscosity. If these parameters vary within the IONPs, then a distribution of $s$ and $D$ values is expected. In such a scenario one may still solve independently Eqs.~(\ref{eq:LammCylZ}) and (\ref{eq:PDEGravity}), due to their linearity, for fixed $s$ and $D$, thus obtaining a solution $c(z,t;s,D)$. The overall solution $c(z,t)$ would be expressed by means of 
\begin{eqnarray}
c(z,t) = \int_{0}^{\infty}\! \int_{0}^{\infty} f(s,D)\, c(z,t;s,D)\, ds\, dD\, ,
\label{eq:cdistributed}
\end{eqnarray} 
where $f(s,D)$ is a joint probability density for $s$ and $D$ satisfying the condition $\int_{0}^{\infty}\!\! \int_{0}^{\infty}\! f(s,D) ds dD=1$. We refer to the book~\cite{Schuck} (Chapter 5) for further details on the basic framework where additional experimental raw sedimentation data is available. 
}
\par

\bibliographystyle{unsrt}
\bibliography{AMM_78_98_2020_Bibliography}

\end{document}